\documentclass[sn-basic]{sn-jnl}
\jyear{2022}%

\theoremstyle{thmstyleone}%
\theoremstyle{thmstyletwo}%

\theoremstyle{thmstylethree}%

\usepackage{graphicx}
\usepackage{enumerate}
\usepackage{natbib}
\usepackage{url}
\usepackage{color}
\usepackage{ulem}
\usepackage{bm}
\usepackage{multicol}
\usepackage{blindtext}

\usepackage[T1]{fontenc} % makes sure we use T1 font
\usepackage[utf8]{inputenc} % should allow a's and o's with two dots in the text without latex coding
\usepackage[english]{babel}
\usepackage{amsmath,amssymb}
\usepackage{mathtools}

\usepackage{microtype} % makes nicer layout of or rows and letters
\usepackage{rotating} % rotating tables
\usepackage{booktabs}  % nicer looking tables (toprule/bottomrule/etc. in Tables)
\usepackage{array}     % fixing of some spacing issues
\usepackage{multirow} % multiple rows in tables
\usepackage{enumitem}

\usepackage{xcolor} % jen kvuli nasim barevnym komentarum
\usepackage{hyperref}

\newcommand{\pkg}[1]{{\normalfont\fontseries{b}\selectfont #1}}

\usepackage{setspace}
\begin{document}

\title[Anisotropic clusters]{Bayesian Inference for Neyman-Scott Point Processes with Anisotropic Clusters}


\author*[1]{\fnm{Ji\v r\' i} \sur{Dvo\v r\' ak}}\email{dvorak@karlin.mff.cuni.cz}

\author[2]{\fnm{Emily} \sur{Ewers}}\email{emily.ewers@rptu.de}

\author[3]{\fnm{Tom\' a\v s} \sur{Mrkvi\v cka}}\email{mrkvicka.toma@gmail.com}

\author[2]{\fnm{Claudia} \sur{Redenbach}}\email{claudia.redenbach@rptu.de}


\affil*[1]{\orgdiv{Faculty of Mathematics and Physics}, \orgname{Charles University}, \orgaddress{\street{Sokolovsk\'{a} 83}, \city{Prague}, \postcode{186 75}, \country{Czech Republic}}}

\affil [2]{\orgdiv{Mathematics Department}, \orgname{RPTU University Kaiserslautern-Landau}, \orgaddress{\street{Gottlieb-Daimler-Stra{\ss}e}, \city{Kaiserslautern}, \postcode{67663}, \country{Germany}}}

\affil[3]{\orgdiv{Faculty of Agriculture and Technology}, \orgname{University of South Bohemia}, \orgaddress{\street{Studentsk\'a 1668}, \city{\v{C}esk\'e Bud\v{e}jovice}, \postcode{370 01}, \country{Czech Republic}}}


\abstract{
    There are few inference methods available to accommodate covariate-dependent aniso\-tropy in point process models. To address this, we propose an extended Bayesian MCMC approach for Neyman-Scott cluster processes. We focus on anisotropy and inhomogeneity in the offspring distribution. Our approach provides parameter estimates as well as sig\-ni\-fi\-cance tests for the covariates and anisotropy through credible intervals, which are determined by the posterior distributions. Additionally, it is possible to test the hypothesis of constant orientation of clusters or constant elongation of clusters. We demonstrate the applicability of this approach through a simulation study for a Thomas-type cluster process.% and an application to real-world data.
}




\maketitle

\leftline{Declarations of interest: none} 
%\bigskip
%\leftline{Data availability: data used in this manuscript can be downloaded at} %\leftline{https://nehody.cdv.cz/} 
\bigskip

\section{Introduction}\label{sec:intro}

%\claudia{Comment: I would consider Neyman-Scott processes to be a whole family of point process models rather than a specific model. Hence, speaking of "the Neyman-Scott point process" feels a bit strange for me.}

Neyman-Scott point processes \citep{NeymanScott1958} are a family of cluster point process models widely used in fields such as biology, astronomy, forestry, and medicine. The models are constructed in two stages: first, cluster centers are randomly sampled from a Poisson point process; second, conditioned on the locations of these cluster centers, a random number of offspring points are independently and randomly placed around each center. Consequently, stationary Neyman-Scott point process models are defined by three components: the intensity of the Poisson process of cluster centers, the distribution of the number of points per cluster (referred to as cluster size), and the distribution of the displacement of the offspring points relative to their respective cluster centers (referred to as cluster spread).

The most common Neyman-Scott point process model is the (modified) Thomas process \citep{Thomas1949}, where the cluster size follows a Poisson distribution and the cluster spread is governed by a radially symmetric Gaussian distribution. The stationary Thomas process is then characterized by the following parameters: the intensity $\kappa$ of the cluster center process, the mean number of points in a cluster $\alpha$, and the standard deviation $\sigma$ of the Gaussian distribution that determines the spread of the cluster.

Anisotropy can be introduced in point patterns in various ways. In practice, anisotropy has to be detected before deciding for a suitable anisotropic point process model. Nonparametric approaches for detecting anisotropy in a point pattern are summarized in \cite{AnisotropyReview}, while isotropy tests are discussed in \cite{fend2024nonparametricisotropytestspatial,RAJALA2022100716,Pypkowski_2025,wong_isotropy_2016}. In most cases, directed versions of established second-order or nearest neighbour summary statistics are compared for a selected set of directions. 
As parametric models, anisotropic Gibbs point processes are considered in \cite{HABEL2017306}.
Another widely used parametric modeling approach for anisotropic point processes is geometric anisotropy. That is, anisotropy is introduced by a linear transformation of an initially isotropic point process model. This approach can be applied to both regular and clustered models. \citet{MollerToftaker2014} propose moment-based methods for parameter estimation as well as a
Bayesian approach for shot-noise Cox processes, where the intensity function is assumed to be constant or a nonparametric kernel estimate of the intensity function is used in the Bayesian analysis. A non-parametric approach for the estimation of the linear transformation introducing geometric anisotropy is introduced in \cite{RAJALA2016100}. An isotropic parametric model can then be fitted to the point pattern obtained by applying the inverse of the estimated transformation to the data.

In cluster point process models, anisotropy can be introduced in two ways. First, the cluster centers can form an anisotropic point process, i.e., the first stage is anisotropic. Second, the distribution of the displacement of the offspring points, i.e., the second stage, can be anisotropic. In this paper, we will be interested only in second-stage anisotropy and only in planar point processes. A second stage anisotropic Neyman-Scott point process is obtained by allowing for an ellipsoidal cluster shape that is rotated in space. Thus, the anisotropic Thomas process is characterized by two additional parameters: $\sigma$ is replaced by $\sigma_1$ and $\sigma_2$, determining the standard deviations in the direction of the two axes of the ellipse, and the angle $\theta$, determining the orientation of the ellipse.   

In general, inference for stationary Neyman-Scott point processes can be made via minimum contrast, composite likelihood, Palm likelihood, or Bayesian MCMC, see \cite{spatstatBook} for an overview. 
\citet{PJ2013} presented an estimation method based on Palm likelihood and applied it to a Thomas-type process with anisotropic clusters. Bayesian inference for stationary and isotropic Neyman-Scott point processes was carried out with an MCMC algorithm, e.g., in \cite{GT2012, MW2007, M2014, KM2016}. In this approach, the cluster centers and the model parameters are updated in each step of the MCMC algorithm. After converging to the stationary distribution, samples from the posterior distributions of the model parameters can be obtained. The cluster centers are generally viewed as nuisance parameters. Bayesian inference for inhomogeneous  Neyman-Scott point processes with isotropic clusters was described in \citet{DRBM2022}. The paper considered inhomogeneity in cluster centers as well as in cluster size and cluster spread and was accompanied by the R package \pkg{binspp}. In this paper, the approach will be extended to account for anisotropy.

Here, we will concentrate on the Bayesian framework because it showed the best performance for Neyman-Scott point processes with inhomogeneous cluster centers \citep{MMK2014}. The second reason for preferring the Bayesian framework is the higher flexibility, e.g., it was applied for Neyman-Scott point processes with inhomogeneity both in cluster centers and in the distribution of clusters in \citet{MS2017}. Due to this flexibility of the Bayesian approach, we are even able to add dependence of the anisotropy parameters on spatial covariates to the model. \citet{MollerToftaker2014} also presented a Bayesian framework for anisotropic Neyman-Scott point processes, but they do not consider any dependence on covariates. They modeled the processes under the assumption of second-order intensity-reweighted stationarity, which approximately corresponds to inhomogeneity in cluster size.

Although the Bayesian MCMC procedure is time-consuming, its great benefit is that the significance of all covariates and all parameters can be determined from the estimated posterior distributions by Bayesian significance testing \citep{Hoshino2008} using credible intervals. To the best of our knowledge, such tests are not available for any other method for anisotropic models with anisotropy parameters depending on covariates. The significance test for the anisotropy parameters can also be used as a tool for anisotropy detection in Neyman-Scott point process models with complex structure. Additionally, it allows checking the assumption of constant direction of clusters or constant elongation of clusters.

This paper is organized as follows. First, in Section~\ref{sec:NS_process} we describe the anisotropic Neyman-Scott point process model. The Bayesian MCMC algorithm is described in Section~\ref{sec:inference}. Section \ref{sec:simulations} summarizes the results of the simulation study where we investigate models covering all combinations of stationary/non-stationary processes with isotropic/anisotropic clusters.
%, dependent, and non-stationary settings. Section \ref{DS} shows the methodology on real data example of ..................................
Section~\ref{sec:CD} provides a brief discussion.

\section{Anisotropic Neyman-Scott point processes} \label{sec:NS_process}

A stationary but anisotropic Neyman-Scott point process is constructed in two steps. First, the process of cluster centers $C$ is a stationary Poisson point process with intensity $\kappa > 0$. The Neyman-Scott point process is the superposition $X=\cup_{c\in C} X_c$ of clusters $X_c$, which are independent Poisson point processes with intensity functions $\alpha k(u-c, \omega ), u \in \mathbb{R}^2$, depending on $c$ and a parameter vector $\omega$.
%$X_c, c \in C$, 
%are independently attached to every cluster center $c$. 
Here, $\alpha$ is the expected number of offspring points in the cluster and $k( \cdot , \omega)$ is the probability density function that governs the relative displacement of the offspring points around the parent point. Due to $X_c$ being Poisson processes, the distribution of the number of points in a cluster is Poisson with parameter $\alpha$.

The function $k( \cdot, \omega)$ can be directly defined as a bivariate probability density function; in the Gaussian case, this could be done through the covariance matrix, assuming zero mean of the distribution. Since we focus on the interpretation of the model and its parameters, especially in the inhomogeneous case below, we choose to parameterize the covariance matrix $\Sigma$ by rotating a template matrix. Specifically, the template matrix
\begin{align*}
  \Sigma_0 = \begin{pmatrix} \sigma_x^2 & 0 \\ 0 & \sigma_y^2 \end{pmatrix}  
\end{align*}
is diagonal with the parameters $\sigma_x > 0$ and $\sigma_y > 0$ giving the standard deviations in the $x$- and $y$-directions, respectively. Let $R$ denote the transformation matrix corresponding to the counterclockwise rotation by the angle $\theta$, expressed in radians. The covariance matrix is then $\Sigma = R \Sigma_0 R^T$. Altogether, $\omega = (\sigma_x, \sigma_y, \theta)^T$ in this (stationary, anisotropic) case. Note that this approach corresponds to the geometric anisotropy of \citet{MollerToftaker2014} and \citet{RAJALA2016100}.

Inhomogeneity (non-stationarity) can be introduced to a Neyman-Scott point process through various model components. In particular, the parameters $\kappa$, $\alpha$, and $\omega$ controlling the intensity of cluster centers, cluster size, and cluster spread, respectively, can be location-dependent, possibly depending on some spatial covariates; see \cite{DRBM2022} for details.

In this paper, we are interested only in inhomogeneity in anisotropy, i.e., in $\omega$. 
Let $Z_1^x, \ldots, Z_{k_x}^x$ denote the spatial covariates influencing $\sigma_x$ (recall that $\sigma_x^2$ is the first diagonal element of the template matrix before rotation). Similarly, let $Z_1^y, \ldots , Z_{k_y}^y$ be the spatial covariates that influence $\sigma_y$, and $Z_1^\theta, \ldots, Z_{k_\theta}^\theta$ the spatial covariates that influence the rotation angle $\theta$. In the following, we assume that the shape and spread of the cluster corresponding to a parent point in location $u$ is then determined by the parameters $\omega = (\sigma_{x,0}, \ldots, \sigma_{x,k_x}, \sigma_{y,0}, \ldots, \sigma_{y,k_y}, \theta_{0}, \ldots, \theta_{k_\theta})^T$ through the formulas
\begin{align*}
    \sigma_x (u) & = \exp\{ \sigma_{x,0} + \sigma_{x,1} Z_1^x(u) + \ldots + \sigma_{x,k_x} Z_{k_x}^x(u) \}, \; u \in \mathbb{R}^2, \\ 
    \sigma_y (u) & = \exp\{ \sigma_{y,0} + \sigma_{y,1} Z_1^y(u) + \ldots + \sigma_{y,k_y} Z_{k_y}^y(u) \}, \; u \in \mathbb{R}^2, \\
    \theta (u) & = \left[ \theta_0 + \pi \tanh \left( \theta_1 Z_1^\theta(u) + \ldots + \theta_{k_\theta} Z_{k_\theta}^\theta(u) \right) \right] \mod \pi, \; u \in \mathbb{R}^2.
\end{align*}

\begin{figure}
    \centering
    \includegraphics[width=\linewidth]{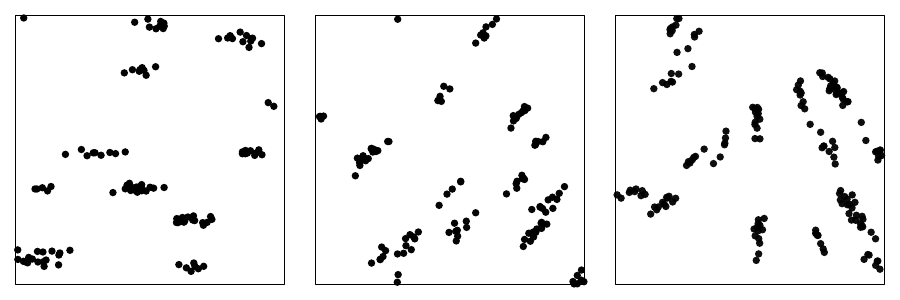}
    \caption{Sample realizations of anisotropic models: clusters aligned with the coordinate axes ($\theta = 0$, left panel), clusters with another orientation ($\theta = \pi/4$, middle panel), clusters with varying orientation ($\theta_0=0, \theta_1 = 1, Z_1^\theta(x,y)=x$).}
    \label{fig:sample_realizations}
\end{figure}

The parametrization is illustrated in Figure~\ref{fig:sample_realizations}. Three realizations from different models were simulated in the unit square window, with $\kappa = 10, \alpha = 10, \sigma_x = 0.02/0.5 = 0.04$ and $\sigma_y = 0.02\cdot 0.5 = 0.01$, motivated by Experiment~3 in Section~\ref{sec:simulations}. The spatial arrangement of points in a cluster is determined by the bivariate Gaussian distribution with a diagonal covariance matrix in the left panel (the covariance matrix $\Sigma$ is the template matrix $\Sigma_0$), with a non-diagonal covariance matrix in the middle panel ($\Sigma = R(\theta) \Sigma_0 R(\theta)^T$, with a fixed rotation matrix $R(\theta)$ corresponding to a non-zero angle $\theta$), and with covariance matrix dependent on the location $u$ of the parent point in the right panel ($\Sigma(u) = R(\theta(u)) \Sigma_0 R(\theta(u))^T$, with the rotation angle $\theta(u)$ depending on a covariate, in this case given by the value of the $x-$coordinate of the parent point).

\section{Bayesian inference} \label{sec:inference}

For Neyman-Scott point processes with anisotropic clusters, we propose the Bayesian approach to inference, as described below. It is based on an MCMC algorithm which consists of alternating updates of the process of cluster centers $C$ and updates of the parameters $\alpha$ and $\omega$. For updating $C$, we use the birth-death-move algorithm described in \citet{MollerWaagepetersen2004}. For updating $\alpha$ and $\omega$, we use the Metropolis-Hastings algorithm. The parameter $\kappa$ is updated based on its relationship with $\alpha$ and the number of observed points as described below. 

To define the posterior distribution of the model parameters, let $C$ denote the homogeneous Poisson point process of cluster centers with intensity $\kappa$, defined on an extended observation window $W_{ext} \supset W$, where $W$ is the window where the offspring points are observed. The reason for considering $W_{ext}$ is that offspring points with parents outside of $W$ may occur in $W$. Let $p(C \mid \kappa)$ denote the probability density function of $C$ with respect to the unit-rate Poisson point process on $W_{ext}$, i.e.
\begin{align*}
    p(C \mid \kappa) = \exp \{ \lvert W_{ext} \rvert - \kappa \lvert W_{ext} \rvert \} \, \kappa^{\lvert C \rvert},
\end{align*}
where $\lvert C \rvert$ denotes the number of points in the configuration $C$ and $\lvert W_{ext} \rvert$ is the area of $W_{ext}$.

Furthermore, let  $p(X \mid C, \kappa, \alpha, \omega)$ denote the probability density function of the point process $X$ conditional on $C, \kappa, \alpha$ and $\omega$, with respect to the unit-rate Poisson process on $W$. Note that conditionally on $C$, $X$ is an inhomogeneous Poisson process with the intensity function $\lambda(\cdot ; C, \alpha, \omega)$ which follows the formula
\begin{align*}
    \lambda(u; C, \alpha, \omega) = \sum_{c \in C} \alpha k(u-c, \omega), \; u \in W,
\end{align*}
where $k$ is the probability density function considered in Section~\ref{sec:NS_process}. It follows that
\begin{align*}
    p(X \mid C, \kappa, \alpha, \omega) = \exp \left\{ \lvert W \rvert - \int_W \lambda(u; C, \alpha, \omega) \, \mathrm{d} u \right\} \prod_{x_i \in X} \lambda(x_i; C, \alpha, \omega).
\end{align*}
The joint posterior distribution of the process $C$ and the parameters is then
\begin{align}
  p(C, \kappa, \alpha, \omega \mid X) \propto p(X \mid C, \kappa, \alpha, \omega) p(C \mid \kappa) p(\alpha) p(\omega) ,
\end{align} 
where $p(\alpha)$ and $p(\omega)$ denote the prior probability density functions for the respective parameters. Note that $\omega$ is a vector of parameters, so the prior must be specified for each component. No prior for $\kappa$ is required because we use a modified Bayesian procedure that was introduced in \citet{KM2016} for the stationary case. It is based on the fact that the expected number of observed points is given by $\mathbb E M = \alpha \kappa \vert W \vert$, where $\vert W \vert$ is the area of the observation window $W$. Thus, $\kappa$ can be determined from $\alpha$ and the natural estimator of $\mathbb E M $, i.e., the observed number of points. The following arguments show that this modification can be used directly in the non-stationary anisotropic case as well.

For non-stationary processes where $\sigma_x(u), \sigma_y(u)$ or $\theta(u)$ are not constant, the intensity function of the offspring process is not constant. Following \cite{MMK2014}, it is given by
\begin{align*}
    \rho(u) = \alpha \kappa \int_{\mathbb{R}^2} k(u - c; \sigma_x(c), \sigma_y(c), \theta(c)) \, \mathrm{d}c, \; u \in \mathbb{R}^2.
\end{align*}
The integral in the formula above cannot be expected to be equal to 1, as the integration is performed with respect to the position of the cluster center $c$. Still, the total number of offspring points in the observation window $W$ will be close to $\alpha \kappa \vert W \vert$:
\begin{align*}
    \int_W \rho(u) \, \mathrm{d} u = \alpha \kappa \int_{\mathbb{R}^2} \int_W k(u - c; \sigma_x(c), \sigma_y(c), \theta(c)) \, \mathrm{d}u \, \mathrm{d}c \approx \alpha \kappa \vert W \vert.
\end{align*}
Replacing the integral on the left-hand side by the observed number of points $n$, this means that we can still take advantage of the link between $\alpha$ and $\kappa$ and estimate only one of these parameters during the run of the MCMC algorithm.

In the simulation studies in Section~\ref{sec:simulations}, the initial configuration of cluster centers is generated randomly from a Poisson process with the intensity function equal to the kernel estimate of the intensity function of the observed offspring pattern $X$. The cluster centers are allowed to occur in the extended observation window $W_{ext}$ during the run of the algorithm. The initial values of the parameters $\alpha$ and $\omega$ need to be specified by the user or sampled from the prior distribution.

The prior distributions for $\alpha$ and $\omega$ also need to be set by the user, taking advantage of any available expert knowledge where possible or choosing uninformative priors. To minimize the effect of the priors on estimation and testing, wide uniform priors seem to be a reasonable choice. The proposal distributions and their standard deviations also need to be chosen, which often requires some preliminary tuning using several shorter runs of the Markov chain in order to achieve reasonable mixing properties. Finally, the desired number of steps, the burn-in period, and the sampling frequency are also determined by the user, usually after several trial runs and after inspecting the relevant diagnostic plots.

\subsection{Parameter estimation}

Inference for the parameters $\alpha$ and $\omega$ is performed based on the estimated posterior distributions. These are obtained from the MCMC samples after appropriate burn-in. Note that it is also possible to make inference about the parameter $\kappa$ using samples from the posterior distribution of $\alpha$ and the link between $\alpha$ and $\kappa$ described in Section~\ref{sec:inference}. However, we do not consider this in the simulation experiments in Section~\ref{sec:simulations} in order to keep the exposition more concise.

As point estimators of the scalar parameters, we suggest using the medians of the samples from the corresponding posterior distributions. For $\theta_0$, the estimation must respect the directional nature of the parameter, and we use the function \texttt{median.circular} from the R package \pkg{circular} in the simulation experiments below.

Furthermore, to quantify the uncertainty in the estimation, we suggest using the credible intervals, defined by the sample 2.5\% and 97.5\% quantiles of the posterior distributions. Once again, for $\theta_0$ the directional nature of the parameter plays an important role, and we use the function \texttt{quantile.circular} from the R package \pkg{circular} in the simulation experiments.

\subsection{Testing of isotropy of clusters}\label{sec:test_of_isotropy}

In the stationary case, where $\sigma_x, \sigma_y$ and $\theta$ do not depend on the location of the parent point,  we are often interested in testing whether the point process is isotropic. The Bayesian approach allows us to check the isotropy assumption by investigating the posterior distribution of the ratio $\sigma_x/\sigma_y$, which describes the circularity of the clusters. Under isotropy, $\sigma_x/\sigma_y = 1$.

During the run of the Markov chain we take samples from the joint posterior distribution of $\sigma_x$ and $\sigma_y$, and hence we can construct a credible interval for $\sigma_x/\sigma_y$. The test outcome is determined simply by looking if the credible interval covers the value 1 (non-rejection) or if the value 1 lies outside of the interval (rejection). The nominal significance level is determined by which quantiles are used as endpoints of the credible interval. We investigate the properties of this test in Experiments 1 and 2 in Section~\ref{sec:simulations}.

This procedure can be extended easily to the non-stationary case where $\sigma_x$ and $\sigma_y$ are constant but $\theta$ possibly depends on some covariates. In this case, it is still possible to construct the credible interval for the circularity parameter $\sigma_x / \sigma_y$ and perform the test as above, see Experiment~3 in Section~\ref{sec:simulations}. The possible dependence of $\theta$ on the location does not change the interpretation of the test outcomes.

In models where $\sigma_x$ and/or $\sigma_y$ are location-dependent, e.g. in Experiment~4 below, the test of isotropy described above is not applicable directly since the value $\sigma_x(u) / \sigma_y(u)$ is not constant. However, one can define a regular grid of points $\{ u_i \}_{i=1}^N$ in the observation window and calculate the values of $\sigma_x(u_i) / \sigma_y(u_i)$ across these points whenever samples from the Markov chain are recorded. In this way, a circularity surface is defined as a transformation of the model parameters, and a generalization of a credible interval can be computed. To do this, we use the methodology of the global envelope tests \citep{GET_JRSSB_2017} to define a credible envelope. Such a concept was used in \citet{Narisetty2016} under the name \textit{simultaneous confidence band} and also mentioned in \citet{MyllymakiMrkvicka2024} under the name \textit{global confidence band}. If the envelope covers the benchmark function (constant 1), the test does not reject the isotropy of clusters. Otherwise, the test rejects.

\subsection{Testing of constant direction of anisotropy}

In models with $\theta$ depending on a single spatial covariate, the test of the null hypothesis of constant anisotropy direction can be performed in the same way as in Section~\ref{sec:test_of_isotropy}, this time using the credible interval for the parameter $\theta_1$. This is investigated in Experiment~3 in Section~\ref{sec:simulations}. When more than one covariate affects $\theta$ in the model, a similar test can be developed using the concept of data depths, central regions, and multivariate quantiles, see e.g. \citet{Liu2003}. However, we do not pursue this direction in this paper.

\section{Simulation study} \label{sec:simulations}

In the description of the prior distributions used in the experiments below, we use the following notation for ease of exposition: $U(a,b)$ denotes the uniform distribution on the interval $[a,b]$, and $LN(m,v)$ denotes the log-normal distribution with mean $m$ and variance $v$. Note that log-normal distributions are frequently parametrized using the mean and variance of the logarithm of the data. Here, we use mean and variance of the non-transformed data as this allows easier comparison, e.g., of the mean of the distribution with the values of model parameters used for simulation of the realizations.

\subsection{Experiment 1}

In this experiment, we consider the case of anisotropic, stationary models. We investigate the precision of the point estimators (compared to the parameter values used for the simulation of the realizations), coverage rates of the credible intervals and, most importantly, the ability of the proposed procedure to detect deviations from isotropy.

\begin{figure}[t]
    \centering
    \includegraphics[width=1\linewidth]{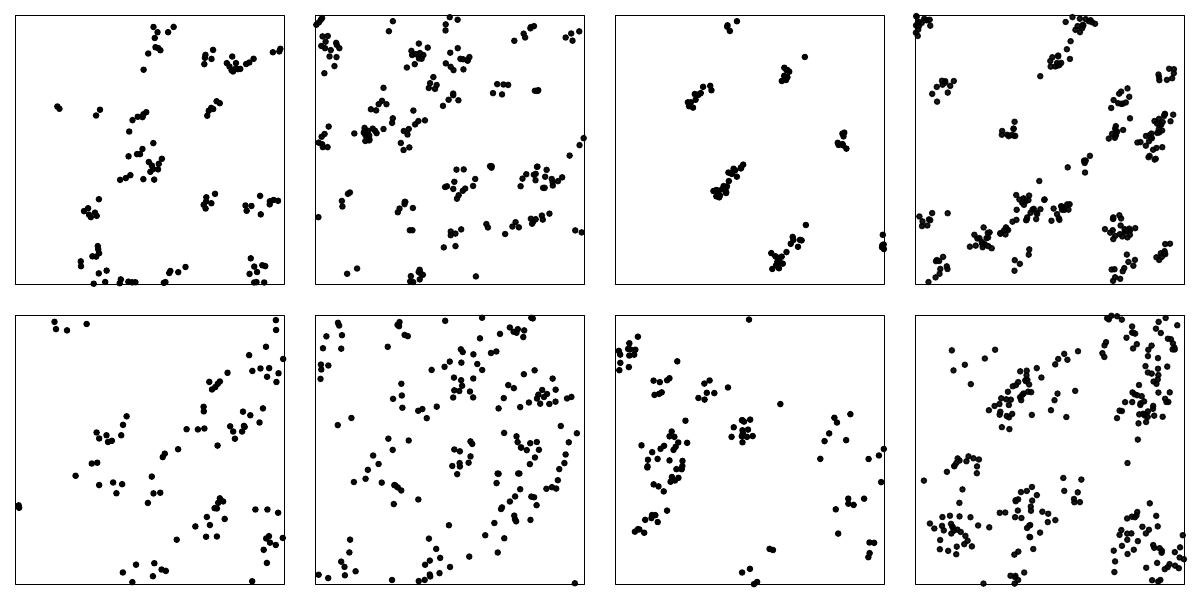}
    \caption{Experiment 1, stationary anisotropic process -- sample realizations from models with strong clustering (top row, models denoted 1--4 in the description of Experiment 1) and weak clustering (bottom row, models denoted 5--8).}
    \label{fig:Exp1_realizations}
\end{figure}

For this experiment, we simulated 20 independent realizations from 8 models exhibiting different behaviors:
weak or strong clustering, with clearly separated clusters or significant cluster overlapping. The total intensity is chosen as $\lambda = 100$ or 200, the mean number of points in a cluster is $\alpha = 5$ or 10 and the clustering parameter is $\sigma = 0.02$ (strong clustering) or 0.04 (weak clustering). To introduce the anisotropy, we define $\sigma_x = \sigma/0.7, \sigma_y = 0.7 \sigma$ and set $\theta_0 = \pi/4$. This results in the value of the circularity parameter $\sigma_x / \sigma_y = 1/0.7^2 \doteq 2.04$. The observation window is $W = [0,1]^2$ and the extended window is $W_{ext} = [-0.2,1.2]^2$. Sample realizations are given in Figure~\ref{fig:Exp1_realizations}.

\begin{figure}[t]
    \centering
    \includegraphics[width=1\linewidth]{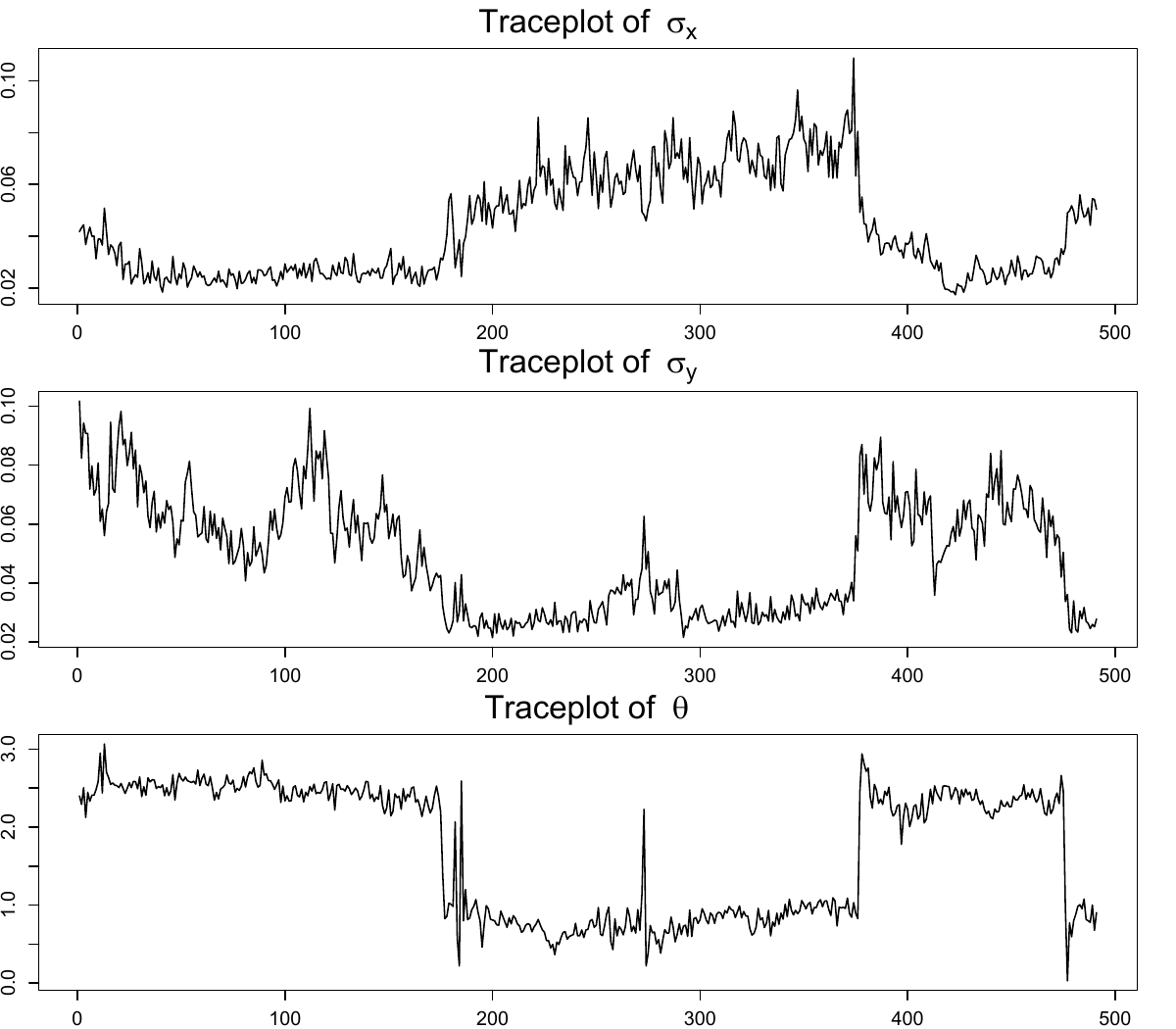}
    \caption{Experiment 1, stationary anisotropic process -- illustration of the fact that the same point process can be described by two different sets of parameters and with very broad uniform priors, the Markov chain may keep switching between the corresponding parts of the state space.}
    \label{fig:switching_roles}
\end{figure}

First, we started the experiments with broad uniform priors for $\alpha, \sigma_x, \sigma_y$ and $\theta_0$. However, this allows the chain to repeatedly switch between the two parts of the state space corresponding to the two sets of parameters describing the same point process, $(\sigma_x, \sigma_y, \theta_0)$ and $(\sigma_y, \sigma_x, \theta_0 + \pi/2)$. This is illustrated in Figure~\ref{fig:switching_roles} where $\sigma_x$ and $\sigma_y$ switch roles several times, together with the corresponding change in the $\theta_0$ values. Naturally, this completely distorts the estimated posterior distributions of $\sigma_x, \sigma_y$ and $\theta_0$ and the corresponding credible intervals. If such behavior was observed in a practical application, one would remedy this issue by selecting different prior distributions (and possibly also different proposal distributions). This signifies the importance of using diagnostic plots in the Bayesian MCMC approach.

After some tuning, we decided to use two sets of prior distributions and present the results for both to allow the reader to assess the influence of the priors.

\subsubsection{Uniform priors}\label{subsubsec:Exp1_uniform}

First, we consider the following uniform priors, the same for all 8 models:
\begin{itemize}
    \item for the parameter $\alpha$, we use the $U(1,30)$ distribution,
    \item for $\theta_0$, we use the $U(0,\pi/2)$ distribution (which alone prevents the problems illustrated in Figure~\ref{fig:switching_roles}),
    \item for $\sigma_x$ and $\sigma_y$, we use the $U(0.002,0.2)$ distribution.
\end{itemize}

Furthermore, the initial state of the Markov chain is the same for all models and all the realizations: $\alpha = 7, \sigma_x = 0.05, \sigma_y = 0.01$, and $\theta_0 = \pi/3$. The proposal distributions for the updates of $\alpha, \sigma_x, \sigma_y$ and $\theta_0$ are centered normal with the standard deviations given in Table~\ref{tab:Exp1_proposal_sds}. For the move update of the cluster centers, the proposal distribution is bivariate normal with zero mean and standard deviation 0.025. Finally, for each realization, the Markov chain is run for 50,000 steps, recording the samples every 100 steps. A burn-in of 25,000 steps is applied before computing the posterior medians and credible intervals for the individual parameters.

\begin{table}[t]
  \renewcommand{\arraystretch}{1.3}

    \centering
    \begin{tabular}{c|cccc}
        Parameter & $\alpha$ & $\sigma_x$ & $\sigma_y$ & $\theta_0$ \\ \hline
        Models 1--4 & 4 & 0.01 & 0.005 & 0.2 \\
        Models 5--8 & 4 & 0.02 & 0.010 & 0.2 \\
    \end{tabular}
    \caption{Experiment 1, stationary anisotropic process -- standard deviations of the proposal distributions for updates of different model parameters.}
    \label{tab:Exp1_proposal_sds}
\end{table}

To describe the performance of the proposed approach, we report the coverage rates of the credible intervals for the different model parameters, see Table~\ref{tab:Exp1_coverage_rates_both}. Note that for the case of uniform priors, the first value in a given column is relevant; the second value gives the coverage rate for the case of informative priors, see below. The coverage rates are computed as the fraction of realizations (out of 20) for which the estimated 95\% credible interval covers the parameter value used for the simulation of the realizations of the given model. This means that the desired coverage rate is 0.95. The table shows that in all models (with the exception of model 6 which is the most difficult to estimate due to severe cluster overlapping), the coverage rates are acceptable, even though the priors are uniform and the priors and the proposal distributions were not tailored individually to each model.

\begin{table}[t]
    \renewcommand{\arraystretch}{1.3}

    \centering
    \begin{tabular}{c|ccc|ccccc}
        & \multicolumn{3}{c|}{Model parameters} & \multicolumn{5}{c}{Coverage rates} \\ \hline
        Model & $\lambda$ & $\alpha$ & $\sigma$ & $\alpha$ & $\sigma_x$ & $\sigma_y$ & $\theta_0$ & $\sigma_x / \sigma_y$ \\ \hline
        1 & 100 & 5  & 0.02 & 0.85/0.85 & 0.75/0.85 & 0.95/0.95 & 0.95/0.95 & 0.80/0.95 \\
        2 & 200 & 5  & 0.02 & 0.90/0.95 & 0.95/0.95 & 0.85/0.90 & 0.80/0.95 & 0.85/0.90 \\
        3 & 100 & 10 & 0.02 & 0.90/0.95 & 0.90/0.90 & 0.90/0.90 & 0.90/0.95 & 0.90/0.95 \\
        4 & 200 & 10 & 0.02 & 1.00/0.95 & 0.95/1.00 & 0.90/0.90 & 1.00/0.95 & 1.00/1.00 \\
        5 & 100 & 5  & 0.04 & 0.90/1.00 & 0.90/1.00 & 0.85/1.00 & 0.90/1.00 & 0.90/1.00 \\
        6 & 200 & 5  & 0.04 & 0.55/0.85 & 0.65/1.00 & 0.85/0.95 & 0.95/1.00 & 0.85/1.00 \\
        7 & 100 & 10 & 0.04 & 0.90/0.95 & 0.95/1.00 & 0.90/0.95 & 0.90/0.85 & 1.00/1.00 \\
        8 & 200 & 10 & 0.04 & 0.85/0.80 & 0.95/0.95 & 0.95/1.00 & 0.90/0.90 & 0.90/1.00 \\ \hline
    \end{tabular}
    \caption{Experiment 1, stationary anisotropic process -- coverage rates of the 95\% credible intervals for different model parameters (columns) and for different models considered in this experiment (rows). The first value in a given column corresponds to the case with \textbf{uniform priors}, the second value corresponds to the case with \textbf{informative priors}.}
    \label{tab:Exp1_coverage_rates_both}
\end{table}

The empirical relative biases and relative mean squared errors of the point estimators are given in Table~\ref{tab:Exp1_numeric_results_uniform} in Appendix~\ref{app:simulations}. In short, the point estimators are rather precise (again, with the exception of model 6), even though the prior distributions were flat.

\begin{figure}[t]
    \centering
    \includegraphics[width=1\linewidth]{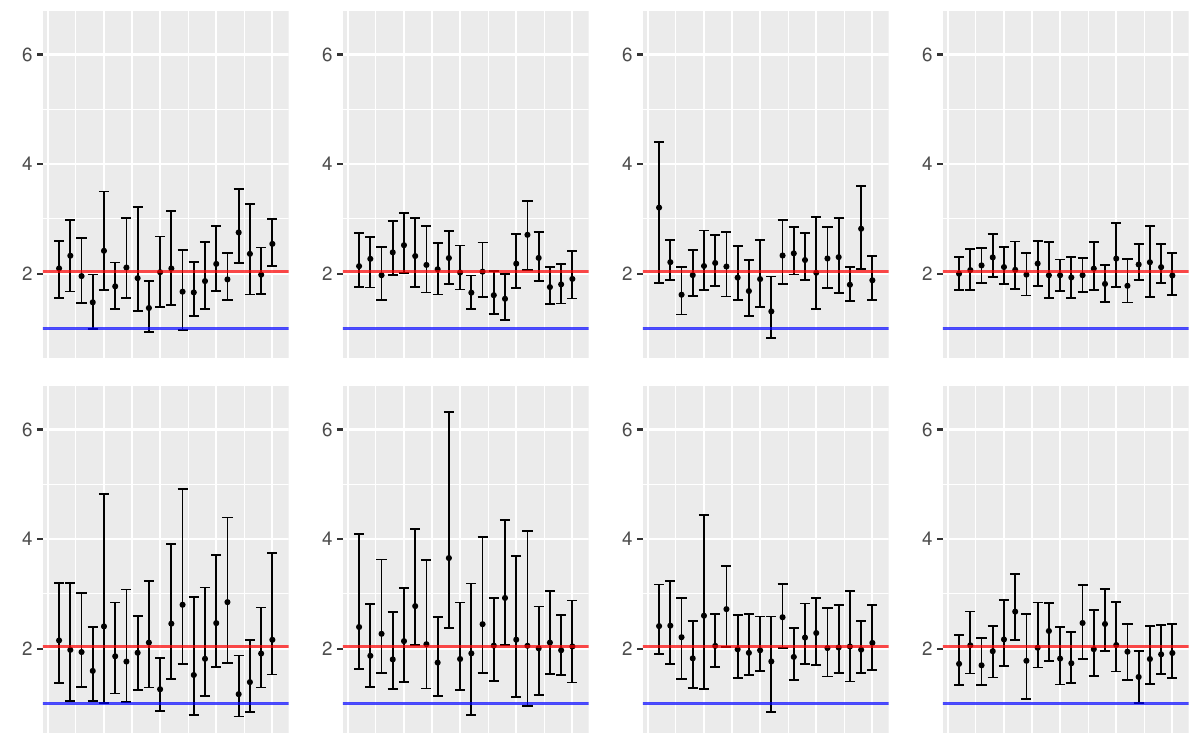}
    \caption{Experiment 1, stationary anisotropic process, \textbf{uniform priors} -- estimated circularity (point estimates with the corresponding 95\% credible intervals) for 20 independent realizations from models 1 to 4 (top row, left to right) and 5 to 8 (bottom row). The red lines show the circularity of the models used for the simulation of the realizations, and the blue lines show the benchmark value 1 corresponding to isotropy.}
    \label{fig:Exp1_circularity_uniform}
\end{figure}

Finally, we assess the ability of the proposed approach to detect deviations from isotropy. To do that, we consider the 95\% credible interval for the circularity parameter $\sigma_x / \sigma_y$ and reject the null hypothesis of isotropy if this interval does not contain the value 1. Figure~\ref{fig:Exp1_circularity_uniform} shows the credible intervals for each realization from the different models considered here. Based on these intervals, we reject the null hypothesis of isotropy for nearly all realizations, with most exceptions corresponding to model 5. This indicates a high power of the test. Note also that the theoretical value 2.04 of the circularity parameter is covered by approx. 90--95 \% of the realizations, see Figure~\ref{fig:Exp1_circularity_uniform} and the last column of Table~\ref{tab:Exp1_coverage_rates_both}.

\subsubsection{Informative priors for $\sigma_x$ and $\sigma_y$}

\begin{figure}
    \centering
    \includegraphics[width=1\linewidth]{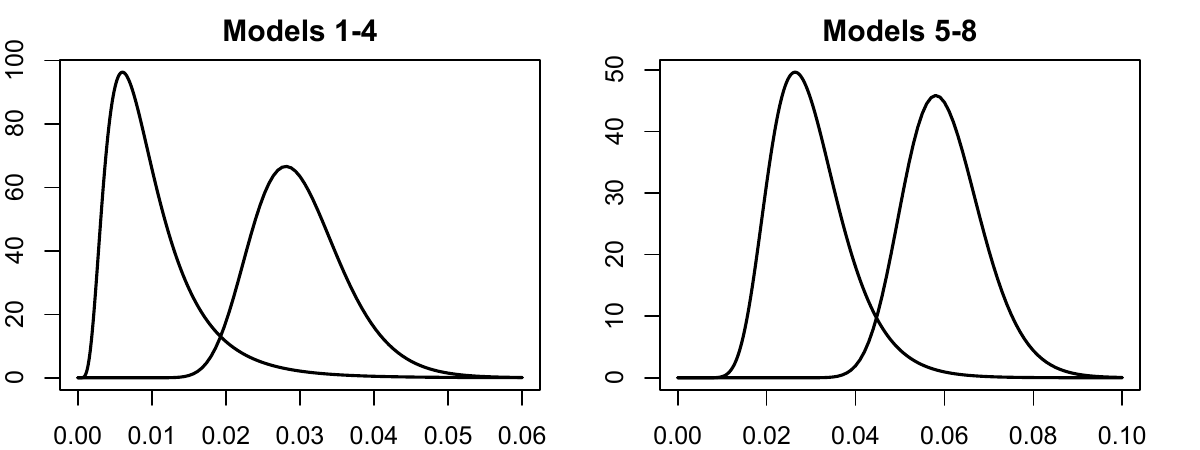}
    \caption{Densities of the prior distributions for $\sigma_x$ and $\sigma_y$ in models 1 to 4 (left) and models 5 to 8 (right).}
    \label{fig:Exp1_priors}
\end{figure}

Second, we use narrower priors for $\sigma_x$ and $\sigma_y$ to illustrate the effect of possible expert knowledge. In this case, we use different priors for $\sigma_x$ and $\sigma_y$ in models with strong clustering ($\sigma = 0.02$, models denoted 1--4 below) and weak clustering, respectively ($\sigma = 0.04$, models denoted 5--8). In particular,
\begin{itemize}
    \item for models 1--4, we use for $\sigma_x$ the $LN(0.03,4\cdot 10^{-5})$ distribution, for $\sigma_y$ we use the $LN(0.01, 4\cdot 10^{-5})$ distribution,
    \item for models 5--8, we use for $\sigma_x$ the $LN(0.06,8\cdot 10^{-5})$, for $\sigma_y$ we use the $LN(0.03, 8\cdot 10^{-5})$ distribution.
\end{itemize}
The densities of the prior distributions for $\sigma_x$ and $\sigma_y$ are given in Figure~\ref{fig:Exp1_priors}. The remaining priors, proposal distributions, initial values etc. remain the same as above in Section~\ref{subsubsec:Exp1_uniform}.

The coverage rates of the credible intervals for the different model parameters are given in Table~\ref{tab:Exp1_coverage_rates_both}. Note that for the case of informative priors, the second value in a given column is relevant; the first value gives the coverage rate for the case of uniform priors, see above. The coverage rates are in most cases closer to the desired value of 0.95, compared to the case of uniform priors. The difference is most prominent in model 6 where the informative priors help the most to stabilize the estimation procedure.

The empirical relative biases and relative mean squared errors of the point estimators are given in Table~\ref{tab:Exp1_numeric_results_informative} in Appendix~\ref{app:simulations}. As expected, the performance is improved compared to the case of uniform priors, most notably in models 5 and 6 with considerable cluster overlapping and for parameters $\sigma_x$ and $\sigma_y$ for which the informative priors were used.

\begin{figure}[t]
    \centering
    \includegraphics[width=1\linewidth]{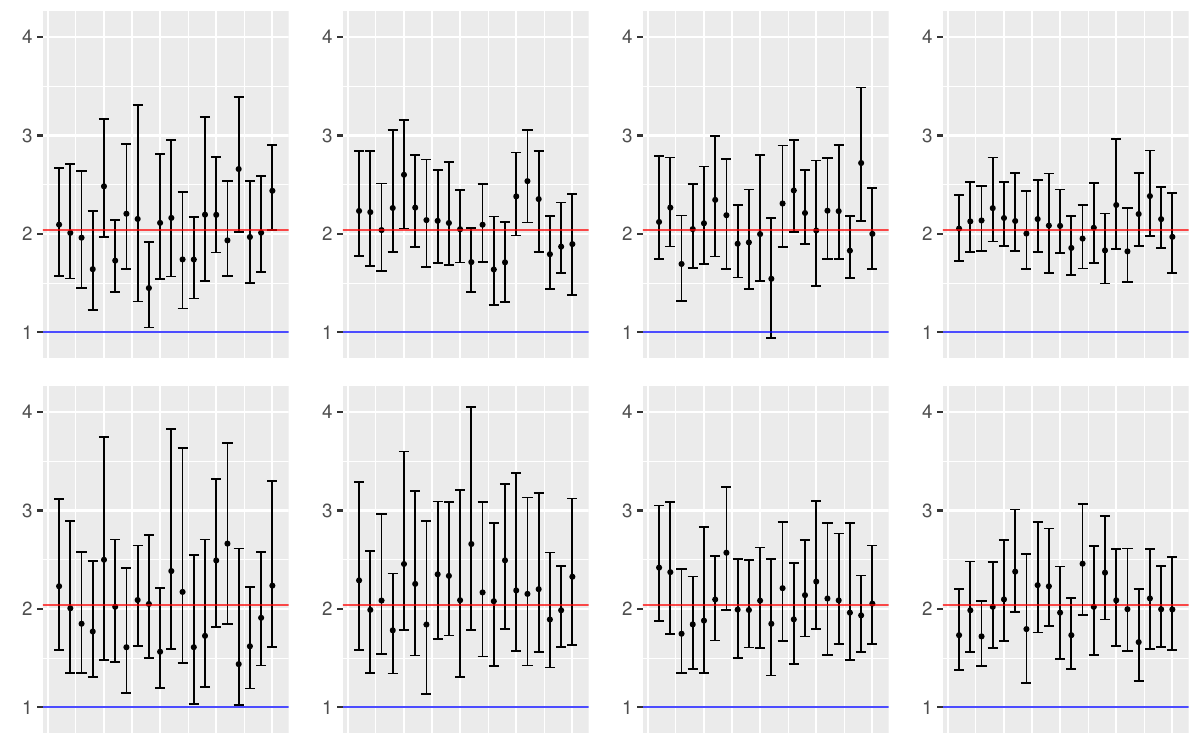}
    \caption{Experiment 1, stationary anisotropic process, \textbf{informative priors} -- estimated circularity (point estimates with the corresponding 95\% credible intervals) for 20 independent realizations from models 1 to 4 (top row, left to right) and 5 to 8 (bottom row). The red lines show the circularity of the models used for the simulation of the realizations, and the blue lines show the benchmark value 1 corresponding to isotropy.}
    \label{fig:Exp1_circularity_informative}
\end{figure}

To assess the ability of the proposed approach to detect deviations from isotropy, we again consider the 95\% credible interval for the circularity parameter $\sigma_x / \sigma_y$. Figure~\ref{fig:Exp1_circularity_informative} shows the credible intervals for each realization from the different models considered here. Based on these intervals, we reject the null hypothesis of isotropy for all realizations except for one realization from model~3, indicating a high power of the test. Note also that the theoretical value 2.04 of the circularity parameter is covered by approx. 95 \% of the realizations, see Figure~\ref{fig:Exp1_circularity_informative}.

\subsection{Experiment 2 (uniform priors)}

This experiment investigates the properties of the proposed approach in the case of isotropic cluster point processes. The primary goal is to see if the null hypothesis of isotropy is (correctly) not rejected by the test based on the credible interval for the circularity parameter. To achieve this, we run the same version of the MCMC algorithm as in Experiment~1, assuming that possibly $\sigma_x \neq \sigma_y$ and some rotation by $\theta_0$ may apply, while in fact the realizations are simulated with $\sigma_x = \sigma_y = \sigma$. The secondary goal is to assess the coverage rates of the credible intervals for the model parameters and the precision of the point estimators. We remark that the results for the rotation parameter $\theta_0$ are not reported in this experiment since no rotation was applied when simulating the realizations.

When choosing the combinations of parameters for simulating the realizations, we use the same values as in Experiment~1, except that $\sigma_x = \sigma_y = \sigma$ in this isotropic case. We simulated 20 independent realizations from the eight resulting models and applied the proposed procedure to them. The same uniform priors, proposal distributions, and initial values as in Experiment~1 were used, see Section~\ref{subsubsec:Exp1_uniform}.

\begin{figure}[t]
    \centering
    \includegraphics[width=1\linewidth]{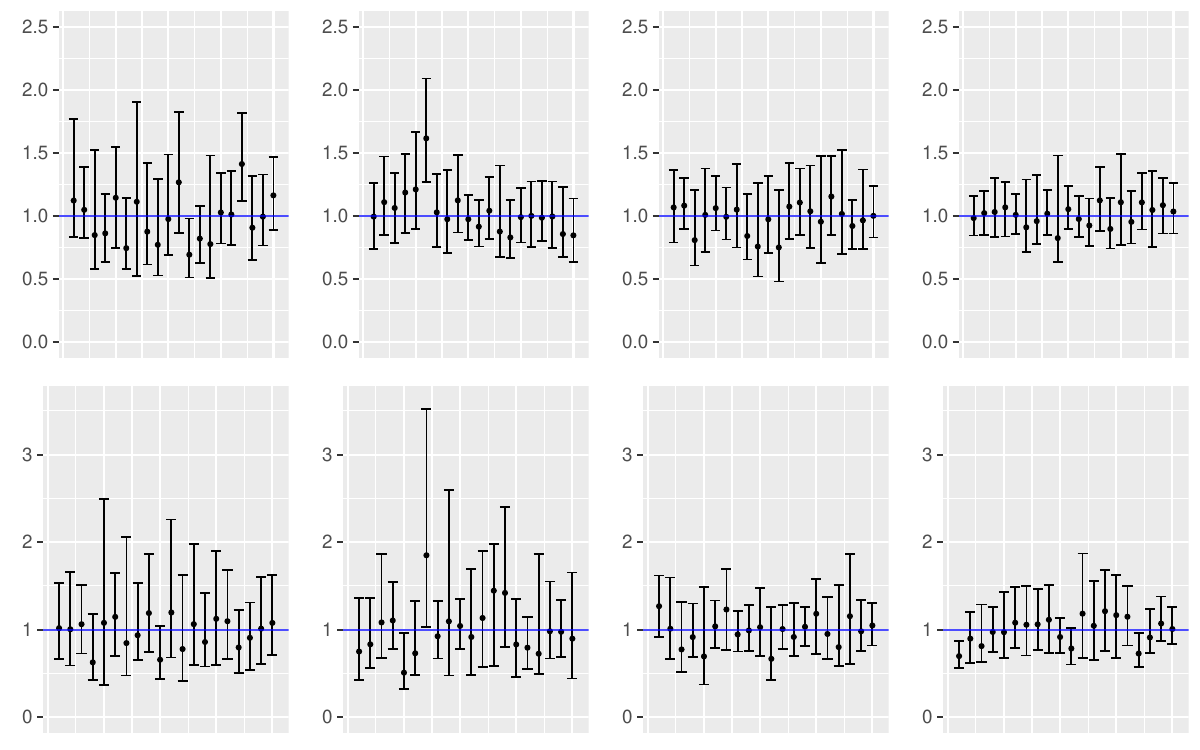}
    \caption{Experiment 2, stationary isotropic process, \textbf{uniform priors} -- estimated circularity (point estimates with the corresponding 95\% credible intervals) for 20 independent realizations from models 1 to 4 (top row, left to right) and 5 to 8 (bottom row). The blue lines show the benchmark value 1 corresponding to isotropy. This value was used to generate the realizations in this experiment.}
    \label{fig:Exp2_circularity_uniform}
\end{figure}

We first assess the credible intervals for the circularity parameter, see Figure~\ref{fig:Exp2_circularity_uniform} and the last column of Table~\ref{tab:Exp2_coverage_rates_uniform}. In this experiment, the null hypothesis of isotropy holds, so it is desired that the 95\% credible intervals avoid the benchmark value 1 (i.e. the test incorrectly rejects the null hypothesis) in 5~\% of cases. Figure~\ref{fig:Exp2_circularity_uniform} shows that the fraction of realizations for which the null hypothesis is (mistakenly) rejected is, for models 1 to 8, equal to 0.10, 0.05, 0.00, 0.00, 0.00, 0.10, 0.00, and 0.10, respectively. Note that these values can be also determined from the last column of Table~\ref{tab:Exp2_coverage_rates_uniform} as the complement to 1. We interpret these results as a reasonable performance of the isotropy test.

The coverage rates are reported in Table~\ref{tab:Exp2_coverage_rates_uniform}. They show satisfactory performance, with the exception of the parameter $\alpha$ in model 6. However, this model is the most challenging to estimate, with the most prominent cluster overlapping (a high number of clusters with a larger spread).

\begin{table}[t]
    \renewcommand{\arraystretch}{1.3}

    \centering
    \begin{tabular}{c|ccc|cccc}
        & \multicolumn{3}{c|}{Model parameters} & \multicolumn{4}{c}{Coverage rates} \\ \hline
        Model & $\lambda$ & $\alpha$ & $\sigma$ & $\alpha$ & $\sigma_x$ & $\sigma_y$ & $\sigma_x / \sigma_y$ \\ \hline
        1 & 100 & 5  & 0.02 & 0.90 & 0.85 & 1.00 & 0.90 \\
        2 & 200 & 5  & 0.02 & 0.90 & 0.95 & 0.95 & 0.95 \\
        3 & 100 & 10 & 0.02 & 0.90 & 0.95 & 1.00 & 1.00 \\
        4 & 200 & 10 & 0.02 & 1.00 & 0.95 & 0.95 & 1.00 \\
        5 & 100 & 5  & 0.04 & 0.95 & 1.00 & 0.90 & 1.00 \\
        6 & 200 & 5  & 0.04 & 0.60 & 0.80 & 0.90 & 0.90 \\
        7 & 100 & 10 & 0.04 & 0.95 & 1.00 & 1.00 & 1.00 \\
        8 & 200 & 10 & 0.04 & 0.95 & 1.00 & 0.90 & 0.90 \\ \hline
    \end{tabular}
    \caption{Experiment 2, stationary isotropic process, \textbf{uniform priors} -- coverage rates of the 95\% credible intervals for different model parameters (columns) and for different models considered in this experiment (rows).}
    \label{tab:Exp2_coverage_rates_uniform}
\end{table}

The empirical relative biases and relative mean squared errors of the point estimators are given in Table~\ref{tab:Exp2_numeric_results_uniform} in Appendix~\ref{app:simulations}. Again, the point estimators are rather precise (with the exception of model 6), even though the prior distributions were flat.

\subsection{Experiment 3 (uniform priors)}

Now, we consider the case of non-stationary models with anisotropic clusters. We let the direction of anisotropy depend on a covariate, whereas the strength of anisotropy is constant. For this experiment, we simulated 20 independent realizations from 4 different models. Compared to previous experiments, a~higher number of observed points is needed to capture the more complex structure of the models. Hence, we fix the (approximate) mean number of points in the observation window to be $\lambda = 200$ by setting the intensity of the parent process $\kappa = 20$ or 40 and the mean number of offsprings per parent $\alpha = 10$ or~5, respectively (recall the discussion of the intensity function of non-stationary Neyman-Scott processes in Section~\ref{sec:inference}).

\begin{figure}[t]
    \centering
    \includegraphics[width=1\linewidth]{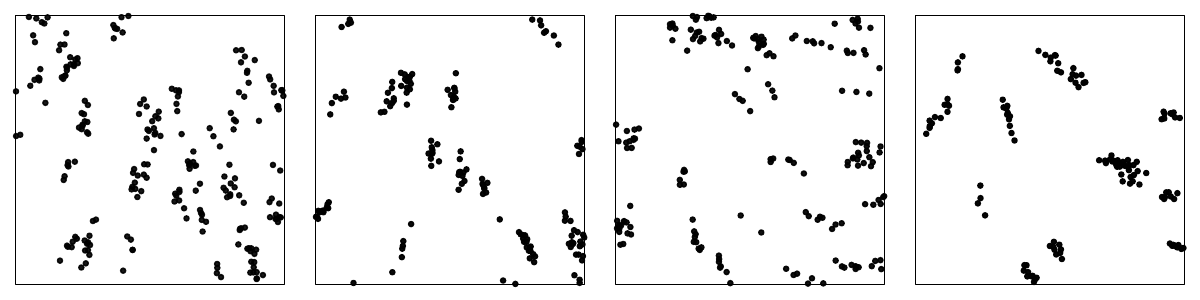}
    \caption{Experiment 3, non-stationary process with anisotropic clusters -- sample realizations from models 1--4 (left to right).}
    \label{fig:Exp3_realizations}
\end{figure}

For the same reason, we consider only stronger clustering here, setting $\sigma = 0.02$, and stronger anisotropy specified by $\sigma_x = \sigma/0.5$ and $\sigma_y = 0.5 \sigma$, resulting in the value of the circularity parameter $\sigma_x / \sigma_y = 1/0.5^2 = 4$. The orientation of a cluster corresponding to a parent point located at $u \in \mathbb{R}^2$ is determined by the rotation angle $\theta (u) = \left[ \theta_0 + \pi \tanh \left( \theta_1 Z_1^\theta(u) \right) \right] \mod \pi$, where $\theta_0 = \pi/4$, $\theta_1 = 0.5$ or 1 and $Z_1^\theta(u) = x, u = (x,y) \in \mathbb{R}^2$. Again, the observation window is $[0,1]^2$, and sample realizations are given in Figure~\ref{fig:Exp3_realizations}.

For conciseness, we consider only uniform priors in this experiment. Specifically, for parameters $\alpha, \sigma_x, \sigma_y$ and $\theta_0$, we use the priors from Section~\ref{subsubsec:Exp1_uniform}. For $\theta_1$, we use the $U[-1,2]$ distribution. Similarly for the initial values of the model parameters -- for $\alpha, \sigma_x, \sigma_y$ and $\theta_0$ we use the values from Section~\ref{subsubsec:Exp1_uniform}. Additionally, for $\theta_1$, we use 0.75 as the starting value. After some tuning, the standard deviations of the proposal distributions were set in the following way: 3 for updates of $\alpha$, 0.01 for $\sigma_x$, 0.002 for $\sigma_y$, 0.1 for $\theta_0$ and $\theta_1$, 0.25 for the move update of the cluster centers. The total number of steps, the sampling frequency and the burn-in are chosen in the same way as in the previous experiments.

We remark that with the choices described above, for 7 out of the 80 realizations in this experiment, the traceplots indicated that the Markov chain makes long jumps between different parts of the state space, similar to the behavior illustrated in Figure~\ref{fig:switching_roles}. This means that in the more complex model considered here, it is not enough to restrict the prior for $\theta_0$ to the interval $(0, \pi/2)$ since the covariate also influences the cluster orientations. Rather than using more informative priors for the sake of the minority of realizations, we decided to run the Markov chain with a different seed for these realizations. With a new seed, the problematic behavior did not appear, and the updated results are reported below. This once again emphasizes the importance of traceplots in the diagnostics of the MCMC runs.

The coverage rates of the credible intervals for the different model parameters are given in Table~\ref{tab:Exp3_coverage_rates}. The table shows that in all models, the coverage rates are close to the desired value of 0.95, even though the priors are uniform and the priors and the proposal distributions were not tailored individually to each model.

The empirical relative biases and relative mean squared errors of the point estimators are given in Table~\ref{tab:Exp3_numeric_results_uniform} in Appendix~\ref{app:simulations}. In short, the point estimators are very precise, even though the prior distributions were flat.

\begin{figure}[t]
    \centering
    \includegraphics[width=1\linewidth]{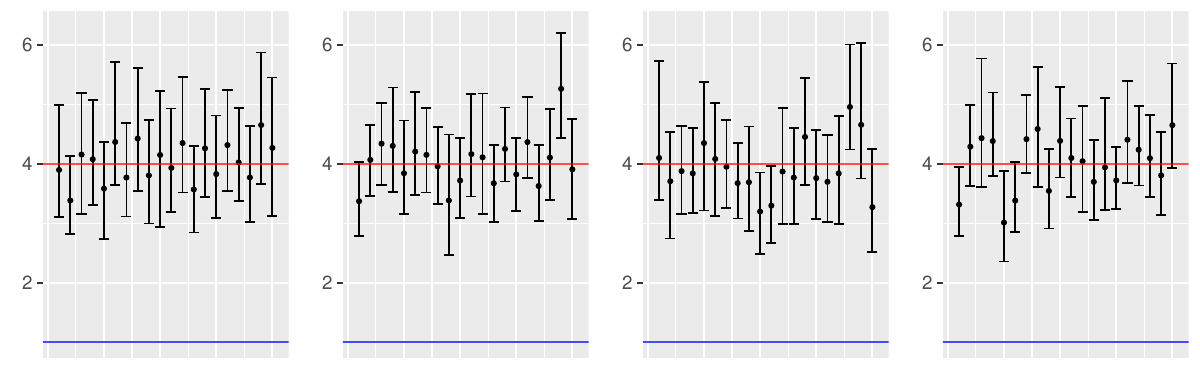}
    \caption{Experiment 3, non-stationary process with anisotropic clusters, \textbf{uniform priors} -- estimated circularity (point estimates with the corresponding 95\% credible intervals) for 20 independent realizations from models 1 to 4 (left to right). The red lines show the circularity of the models used for the simulation of the realizations, and the blue lines show the benchmark value 1 corresponding to isotropy.}
    \label{fig:Exp3_circularity}
\end{figure}

As for the test of isotropy, Figure~\ref{fig:Exp3_circularity} shows the credible intervals for the circularity parameter $\sigma_x / \sigma_y$ for each realization from the different models considered here. Based on these intervals, we reject the null hypothesis of isotropy for all realizations. This indicates a very high power of the test. Note also that the theoretical value 4 of the circularity parameter is covered by approx. 90--95 \% of the realizations, see Figure~\ref{fig:Exp3_circularity} and the last column of Table~\ref{tab:Exp3_coverage_rates}.

\begin{table}[t]
    \renewcommand{\arraystretch}{1.3}

    \centering
    \begin{tabular}{c|cccc|cccccc}
        & \multicolumn{4}{c|}{Model parameters} & \multicolumn{6}{c}{Coverage rates} \\ \hline
        Model & $\lambda$ & $\alpha$ & $\sigma$ & $\theta_1$ & $\alpha$ & $\sigma_x$ & $\sigma_y$ & $\theta_0$ & $\theta_1$ & $\sigma_x / \sigma_y$ \\ \hline
        1 & 200 & 5  & 0.02 & 0.5 & 0.95 & 1.00 & 1.00 & 0.95 & 0.95 & 1.00 \\
        2 & 200 & 10 & 0.02 & 0.5 & 0.90 & 0.95 & 0.90 & 0.95 & 0.90 & 0.95 \\
        3 & 200 & 5  & 0.02 & 1   & 0.85 & 0.95 & 0.85 & 1.00 & 0.95 & 0.85 \\
        4 & 200 & 10 & 0.02 & 1   & 0.95 & 0.90 & 0.90 & 0.85 & 0.95 & 0.90 \\ \hline
    \end{tabular}
    \caption{Experiment 3, non-stationary process with anisotropic clusters, \textbf{uniform priors} -- coverage rates of the 95\% credible intervals for different model parameters (columns) and for different models considered in this experiment (rows).}
    \label{tab:Exp3_coverage_rates}
\end{table}

\subsection{Experiment 4 (uniform priors)}

Finally, we investigate the case of non-stationary models with isotropic clusters. We let the cluster spread depend on a covariate, with all clusters being circular. We simulated 20 independent realizations from 4 different models, again with the (approximate) mean number of points in the observation window fixed to $\lambda = 200$ by setting the intensity of the parent process $\kappa = 20$ or 40 and the mean number of offsprings per parent $\alpha = 10$ or 5, respectively.

\begin{figure}[t]
    \centering
    \includegraphics[width=1\linewidth]{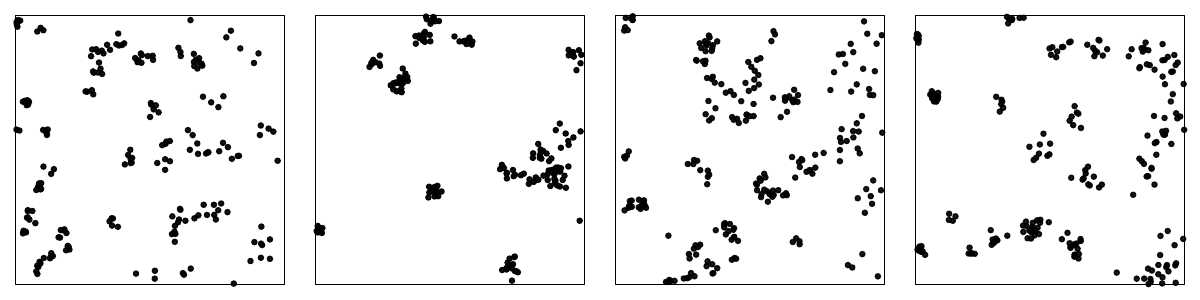}
    \caption{Experiment 4, non-stationary process with isotropic clusters -- sample realizations from models 1--4 (left to right).}
    \label{fig:Exp4_realizations}
\end{figure}

We make all clusters circular by setting $\sigma_x(u) = \sigma_y(u)$, but we allow this function to depend on a covariate through the formula
\begin{align*}
    \sigma_x (u) & = \exp\{ \sigma_{0,x} + \sigma_{1,x} Z_1^x(u) \}, \; u \in \mathbb{R}^2,
\end{align*}
where $Z_1^x(u) = x, u = (x,y) \in \mathbb{R}^2$. The observation window is $[0,1]^2$ and by setting $\sigma_{0,x} = \log(0.01)$ and $\sigma_{1,x} = 1$ or 1.5 we obtain models where clusters are very tight in the left part of the observation window but increase their spread towards the right part of the window. Sample realizations are given in Figure~\ref{fig:Exp4_realizations}.

We fit the model that allows for anisotropic clusters, that is, containing $\sigma_{0,x}, \sigma_{1,x}, \sigma_{0,y}, \sigma_{1,y}$ and $\theta_0$ and the appropriate covariates $Z_1^x(u) = Z_1^y(u)$. Again, we consider only uniform priors in this experiment. Specifically, for parameters $\alpha$ and $\theta_0$, we use the priors from Section~\ref{subsubsec:Exp1_uniform}. For $\sigma_{0,x}$ and $\sigma_{0,y}$ we use the $U[\log(0.002),\log(0.2)]$ distribution and for $\sigma_{1,x}$ and $\sigma_{1,y}$ the $U[-5,5]$ distribution. For the initial values of the model parameters, we use 7 for $\alpha$, $\log(0.015)$ for $\sigma_{0,x}$ and $\sigma_{0,y}$, 1.25 for $\sigma_{1,x}$ and $\sigma_{1,y}$ and 0 for $\theta_0$. The standard deviations of the proposal distributions for updates of the model parameters are inspired by the previous experiments: 3 for the parameter $\alpha$, 0.1 for parameters $\sigma_{0,x}, \sigma_{0,y}, \sigma_{1,x}, \sigma_{1,y}$ and $\theta_0$ (note that in this experiment, $\sigma_{0,x}$ and $\sigma_{0,y}$ operate on the log scale), 0.25 for the move update of the cluster centers. The total number of steps, the sampling frequency and the burn-in are chosen in the same way as in the previous experiments.

We are interested mainly in the ability of the inferential procedure to identify the isotropic nature of the clusters, even though the cluster spread varies across the observation window. For this purpose, we investigate the posterior distribution of the circularity surface $\sigma_x(u) / \sigma_y(u)$ and compute the corresponding credible envelope, see Section~\ref{sec:test_of_isotropy}. If the (constant) benchmark function $f(u) = 1$ is not fully covered by the credible envelope, the hypothesis of isotropic shape of clusters is rejected. The fraction of realizations for which the hypothesis is (mistakenly) rejected is, for models 1 to 4, equal to 0.05, 0.00, 0.00, and 0.20, respectively. Note that these values can be also determined from the last column of Table~\ref{tab:Exp4_coverage_rates} as the complement to 1. We interpret these results as a reasonable performance of the test. Illustration of the test and its interpretation is given in Appendix~\ref{app:credible_envelope}.

%We are interested mainly in the ability of the inferential procedure to identify the isotropic nature of the clusters, even though the cluster spread varies across the observation window. For this purpose, we investigate the posterior distribution of the circularity parameter (defined in this case as the median of $\sigma_x(u) / \sigma_y(u)$ across a fixed regular grid of points in $W$) and the corresponding credible intervals, see Figure~\ref{fig:Exp4_circularity} and the last column of Table~\ref{tab:Exp4_coverage_rates}. In this experiment, the null hypothesis of isotropic clusters holds, so it is desired that the 95\% credible intervals avoid the benchmark value 1 (i.e. the test rejects the null hypothesis) in 5 \% of cases. Figure~\ref{fig:Exp4_circularity} shows that the fraction of realizations for which the null hypothesis is (mistakenly) rejected is, for models 1 to 4, equal to 0.10, 0.05, 0.00, and 0.15, respectively. Note that these values can be also determined from the last column of Table~\ref{tab:Exp4_coverage_rates} as the complement to 1. We interpret these results as a reasonable performance of the test.

The coverage rates of the credible intervals for the different model parameters are given in Table~\ref{tab:Exp4_coverage_rates}. The table shows that in all models, the coverage rates are close to the desired value of 0.95, even though the priors are uniform and the priors and the proposal distributions were not tailored individually to each model.

\begin{table}[t]
    \renewcommand{\arraystretch}{1.3}

    \centering
    \begin{tabular}{c|cccc|cccccc}
        & \multicolumn{4}{c|}{Model parameters} & \multicolumn{6}{c}{Coverage rates} \\ \hline
        Model & $\lambda$ & $\alpha$ & $\sigma_{0,x}$ & $\sigma_{1,x}$ & $\alpha$ & $\sigma_{0,x}$ & $\sigma_{1,x}$ & $\sigma_{0,y}$ & $\sigma_{1,y}$ & circ. \\ \hline
        1 & 200 & 5  & $\log(0.01)$ & 1   & 0.90 & 0.90 & 1.00 & 0.95 & 0.95 & 0.95 \\
        2 & 200 & 10 & $\log(0.01)$ & 1   & 0.95 & 1.00 & 0.95 & 1.00 & 1.00 & 1.00 \\
        3 & 200 & 5  & $\log(0.01)$ & 1.5 & 0.85 & 1.00 & 0.95 & 1.00 & 1.00 & 1.00 \\
        4 & 200 & 10 & $\log(0.01)$ & 1.5 & 0.90 & 0.95 & 0.95 & 0.90 & 0.90 & 0.80 \\
    \end{tabular}
    \caption{Experiment 4, non-stationary process with isotropic clusters, \textbf{uniform priors} -- coverage rates of the 95\% credible intervals for different model parameters (columns) and for different models considered in this experiment (rows).}
    \label{tab:Exp4_coverage_rates}
\end{table}

The empirical relative biases and the relative mean square errors of the point estimators are given in Table~\ref{tab:Exp4_numeric_results_uniform} in Appendix~\ref{app:simulations}. In short, the point estimators are very precise, even though the prior distributions were flat.

\section{Conclusions and discussion} \label{sec:CD}

The methodology proposed in this paper performed well in estimating the parameters of the Neyman-Scott point process with anisotropic clusters (Experiment 1). Clearly, the results are sensitive to the chosen priors. Nevertheless, even with flat priors, acceptable results are achieved. The method exhibited good power and significance level for determining anisotropy using the posterior distribution of the circularity parameter (Experiments 1 and 2).

Experiment 3 indicated that the methodology is also capable of detecting anisotropy in the case where the orientation of the clusters is not constant and depends on a spatial covariate. This is in line with earlier findings about the capability of the Bayesian MCMC algorithm to determine the dependencies of the inhomogeneous Neyman-Scott point process on spatial covariates, both in cluster centers and in cluster size and cluster spread \citep{DRBM2022}. The current findings motivate adding the anisotropic versions of the algorithm to the \pkg{binspp} package \citep{DRBM2022} and extending its versatility with anisotropic models and isotropy tests. The core functions of the \pkg{binspp} package are implemented using \pkg{Rcpp} to speed up the computation. Thus, short runs of the chain, useful for tuning up the hyperparameters of the prior distributions, are finished within minutes on a regular laptop. Long runs, used for the actual estimation, may be completed within a few hours. This makes our implementation easily applicable in practice, without the need to worry about the computational demands.

Careful inspection of the Monte Carlo trace plots is vital in any case where MCMC algorithms are involved. Here, the circular nature of the cluster orientation highlights this even more, due to the possibility of jumps between equivalent parametric states. This can be remedied by decreasing the standard deviations of the proposal distributions for parameter updates. However, in this case, the mixing properties of the Markov chain must be checked with caution.

The presented model and the corresponding inference methodology are very flexible, and straightforward generalization is possible to account for the possible three-dimensional nature of the data, for non-Gaussian displacement kernels, and for a combination of anisotropy and inhomogeneity in cluster size and/or inhomogeneity of the parent process.

%\section*{References}
%\bibliographystyle{model2names}\biboptions{authoryear}
%\bibliographystyle{rss}
%\bibliographystyle{apa}
\bibliography{Aniso_references} 

\begin{thebibliography}{25}
\providecommand{\natexlab}[1]{#1}
\providecommand{\url}[1]{{#1}}
\providecommand{\urlprefix}{URL }
\providecommand{\doi}[1]{\url{https://doi.org/#1}}
\providecommand{\eprint}[2][]{\url{#2}}
 \bibcommenthead

\bibitem[{Baddeley et~al(2015)Baddeley, Rubak, and Turner}]{spatstatBook}
Baddeley A, Rubak E, Turner R (2015) Spatial Point Patterns: Methodology and Applications with R. Chapman \& Hall, New York

\bibitem[{Dvořák et~al(2022)Dvořák, Remeš, Beránek, and Mrkvička}]{DRBM2022}
Dvořák J, Remeš R, Beránek L, et~al (2022) binspp: {A}n {R} package for {B}ayesian inference for {N}eyman-{S}cott point processes with complex inhomogeneity structure. \urlprefix\url{https://arxiv.org/abs/2205.07946}

\bibitem[{Fend and Redenbach(2024)}]{fend2024nonparametricisotropytestspatial}
Fend C, Redenbach C (2024) Nonparametric isotropy test for spatial point processes using random rotations. \urlprefix\url{https://arxiv.org/abs/2404.10594}

\bibitem[{Guttorp and Thorarinsdottir(2012)}]{GT2012}
Guttorp P, Thorarinsdottir TL (2012) Bayesian inference for non-{M}arkovian point processes. In: Porcu E, Montero JM, Schlather M (eds) Advances and Challenges in Space-time Modelling of Natural Events. Springer, p 79--102

\bibitem[{Hoshino(2008)}]{Hoshino2008}
Hoshino T (2008) Bayesian significance testing and multiple comparisons from mcmc outputs. Computational Statistics \& Data Analysis 52(7):3543--3559

\bibitem[{Häbel et~al(2017)Häbel, Rajala, Marucci, Boissier, Schladitz, Redenbach, and Särkkä}]{HABEL2017306}
Häbel H, Rajala T, Marucci M, et~al (2017) A three-dimensional anisotropic point process characterization for pharmaceutical coatings. Spatial Statistics 22:306--320

\bibitem[{Kopeck\'{y} and Mrkvi\v{c}ka(2016)}]{KM2016}
Kopeck\'{y} J, Mrkvi\v{c}ka T (2016) On {B}ayesian estimation for {N}eyman-{S}cott point processes. Applications of Mathematics 61(4):503--514

\bibitem[{Liu(2003)}]{Liu2003}
Liu RY (2003) Data depth: Center-outward ordering of multivariate data and nonparametric multivariate statistics. In: Akritas MG, Politis DN (eds) Recent Advances and Trends in Nonparametric Statistics. JAI, Amsterdam, p 155--167

\bibitem[{M{\o}ller and Toftaker(2014)}]{MollerToftaker2014}
M{\o}ller J, Toftaker H (2014) Geometric anisotropic spatial point pattern analysis and cox processes. Scandinavian Journal of Statistics 41(2):414--435

\bibitem[{M{\o}ller and Waagepetersen(2004)}]{MollerWaagepetersen2004}
M{\o}ller J, Waagepetersen RP (2004) Statistical Inference and Simulation for Spatial Point Processes. Chapman \& {Hall/CRC}

\bibitem[{M{\o}ller and Waagepetersen(2007)}]{MW2007}
M{\o}ller J, Waagepetersen RP (2007) Modern statistics for spatial point processes. Scandinavian Journal of Statistics 34(4):643--684

\bibitem[{Mrkvi{\v c}ka(2014)}]{M2014}
Mrkvi{\v c}ka T (2014) Distinguishing different types of inhomogeneity in {N}eyman-{S}cott point processes. Methodology and Computing in Applied Probability 16:385--395

\bibitem[{Mrkvi{\v c}ka and Soubeyrand(2017)}]{MS2017}
Mrkvi{\v c}ka T, Soubeyrand S (2017) On parameter estimation for doubly inhomogeneous cluster point processes. Spatial Statistics 20:191--205

\bibitem[{Mrkvi{\v c}ka et~al(2014)Mrkvi{\v c}ka, Mu{\v s}ka, and Kube{\v c}ka}]{MMK2014}
Mrkvi{\v c}ka T, Mu{\v s}ka M, Kube{\v c}ka J (2014) Two step estimation for {N}eyman-{S}cott point process with inhomogeneous cluster centers. Statistics and Computing 24:91--100

\bibitem[{Myllym{\"a}ki et~al(2017)Myllym{\"a}ki, Mrkvi{\v c}ka, Grabarnik, Seijo, and Hahn}]{GET_JRSSB_2017}
Myllym{\"a}ki M, Mrkvi{\v c}ka T, Grabarnik P, et~al (2017) Global envelope tests for spatial processes. Journal of the Royal Statistical Society: Series B (Statistical Methodology) 79(2):381--404

\bibitem[{Myllymäki and Mrkvička(2024)}]{MyllymakiMrkvicka2024}
Myllymäki M, Mrkvička T (2024) Get: Global envelopes in r. Journal of Statistical Software 111(3):1–40

\bibitem[{Narisetty and Nair(2016)}]{Narisetty2016}
Narisetty NN, Nair VN (2016) Extremal depth for functional data and applications. Journal of the American Statistical Association 111(516):1705--1714

\bibitem[{Neyman and Scott(1958)}]{NeymanScott1958}
Neyman J, Scott EL (1958) Statistical approach to problems of cosmo\-logy. Journal of the Royal Statistical Society, Series B 20:1--43

\bibitem[{Prokešová and Jensen(2013)}]{PJ2013}
Prokešová M, Jensen E (2013) Asymptotic palm likelihood theory for stationary point processes. Annals of the Institute of Statistical Mathematics 65:387--412

\bibitem[{Pypkowski et~al(2025)Pypkowski, Sykulski, and Martin}]{Pypkowski_2025}
Pypkowski JJ, Sykulski AM, Martin JS (2025) Isotropy testing in spatial point patterns: nonparametric versus parametric replication under misspecification. Spatial Statistics 67:100,898

\bibitem[{Rajala et~al(2016)Rajala, Särkkä, Redenbach, and Sormani}]{RAJALA2016100}
Rajala T, Särkkä A, Redenbach C, et~al (2016) Estimating geometric anisotropy in spatial point patterns. Spatial Statistics 15:100--114

\bibitem[{Rajala et~al(2018)Rajala, Redenbach, Särkkä, and Sormani}]{AnisotropyReview}
Rajala T, Redenbach C, Särkkä A, et~al (2018) A review on anisotropy analysis of spatial point patterns. Spatial Statistics 28:141--168

\bibitem[{Rajala et~al(2022)Rajala, Redenbach, Särkkä, and Sormani}]{RAJALA2022100716}
Rajala T, Redenbach C, Särkkä A, et~al (2022) Tests for isotropy in spatial point patterns – a comparison of statistical indices. Spatial Statistics 52:100,716

\bibitem[{Thomas(1949)}]{Thomas1949}
Thomas M (1949) A generalization of {P}oisson's binomial limit for use in ecology. Biometrika 36:18--25

\bibitem[{Wong and Chiu(2016)}]{wong_isotropy_2016}
Wong KY, Chiu SN (2016) Isotropy test for spatial point processes using stochastic reconstruction. Spatial Statistics 15:56--69

\end{thebibliography}

\newpage

\appendix

\section{Additional simulation results}\label{app:simulations}

In this section we provide the relative biases and relative mean squared errors for point estimators of various model parameters from simulation experiments 1 to 4, see Section~\ref{sec:simulations} for details.

\begin{table}[h]
    \renewcommand{\arraystretch}{1.3}

    \centering
    \begin{tabular}{c|ccc|cccc}
        & \multicolumn{3}{c|}{Model parameters} & \multicolumn{4}{c}{Relative bias} \\ \hline
        Model & $\lambda$ & $\alpha$ & $\sigma$ & $\alpha$ & $\sigma_x$ & $\sigma_y$ & $\theta$ \\ \hline
        1 & 100 & 5  & 0.02 & 0.077 & 0.006 & 0.033 & -0.014 \\
        2 & 200 & 5  & 0.02 & 0.019 & 0.034 & 0.025 & 0.025 \\
        3 & 100 & 10 & 0.02 & 0.047 & 0.065 & 0.037 & 0.020 \\
        4 & 200 & 10 & 0.02 & -0.015 & 0.002 & -0.003 & -0.004 \\
        5 & 100 & 5  & 0.04 & 0.121 & 0.040 & 0.110 & 0.004 \\
        6 & 200 & 5  & 0.04 & 0.289 & 0.161 & 0.060 & 0.038 \\
        7 & 100 & 10 & 0.04 & 0.188 & 0.081 & 0.031 & -0.022 \\
        8 & 200 & 10 & 0.04 & 0.034 & 0.033 & 0.064 & -0.015 \\ \hline
    \end{tabular}

    \vspace{0.2cm}

    \begin{tabular}{c|ccc|cccc}
        & \multicolumn{3}{c|}{Model parameters} & \multicolumn{4}{c}{Relative MSE} \\ \hline
        Model & $\lambda$ & $\alpha$ & $\sigma$ & $\alpha$ & $\sigma_x$ & $\sigma_y$ & $\theta$ \\ \hline
        1 & 100 & 5  & 0.02 & 0.039 & 0.018 & 0.026 & 0.036 \\
        2 & 200 & 5  & 0.02 & 0.013 & 0.013 & 0.009 & 0.017 \\
        3 & 100 & 10 & 0.02 & 0.018 & 0.030 & 0.013 & 0.011 \\
        4 & 200 & 10 & 0.02 & 0.007 & 0.004 & 0.005 & 0.005 \\
        5 & 100 & 5  & 0.04 & 0.092 & 0.038 & 0.067 & 0.033 \\
        6 & 200 & 5  & 0.04 & 0.491 & 0.155 & 0.076 & 0.045 \\
        7 & 100 & 10 & 0.04 & 0.086 & 0.028 & 0.013 & 0.035 \\
        8 & 200 & 10 & 0.04 & 0.060 & 0.015 & 0.013 & 0.014 \\ \hline
    \end{tabular}

    \caption{Experiment 1, stationary anisotropic process, \textbf{uniform priors} -- estimated relative bias (top part) and relative mean squared error (bottom part) for different model parameters (columns) and for different models considered in this experiment (rows).}
    \label{tab:Exp1_numeric_results_uniform}
\end{table}

\begin{table}[t]
    \renewcommand{\arraystretch}{1.3}

    \centering
    \begin{tabular}{c|ccc|cccc}
        & \multicolumn{3}{c|}{Model parameters} & \multicolumn{4}{c}{Relative bias} \\ \hline
        Model & $\lambda$ & $\alpha$ & $\sigma$ & $\alpha$ & $\sigma_x$ & $\sigma_y$ & $\theta$ \\ \hline
        1 & 100 & 5  & 0.02 & 0.077 & 0.001 & 0.016 & -0.017 \\
        2 & 200 & 5  & 0.02 & -0.017 & 0.014 & -0.007 & 0.008 \\
        3 & 100 & 10 & 0.02 & 0.022 & 0.028 & 0.003 & 0.017 \\
        4 & 200 & 10 & 0.02 & -0.014 & 0.004 & -0.014 & -0.007 \\
        5 & 100 & 5  & 0.04 & 0.076 & 0.014 & 0.068 & -0.010 \\
        6 & 200 & 5  & 0.04 & 0.091 & 0.048 & -0.006 & -0.007 \\
        7 & 100 & 10 & 0.04 & 0.101 & 0.021 & 0.012 & -0.032 \\
        8 & 200 & 10 & 0.04 & -0.019 & 0.015 & 0.025 & -0.021 \\ \hline
    \end{tabular}

    \vspace{0.2cm}

    \begin{tabular}{c|ccc|cccc}
        & \multicolumn{3}{c|}{Model parameters} & \multicolumn{4}{c}{Relative MSE} \\ \hline
        Model & $\lambda$ & $\alpha$ & $\sigma$ & $\alpha$ & $\sigma_x$ & $\sigma_y$ & $\theta$ \\ \hline
        1 & 100 & 5  & 0.02 & 0.041 & 0.011 & 0.023 & 0.031 \\
        2 & 200 & 5  & 0.02 & 0.012 & 0.005 & 0.009 & 0.019 \\
        3 & 100 & 10 & 0.02 & 0.014 & 0.010 & 0.009 & 0.010 \\
        4 & 200 & 10 & 0.02 & 0.005 & 0.003 & 0.005 & 0.005 \\
        5 & 100 & 5  & 0.04 & 0.041 & 0.003 & 0.032 & 0.029 \\
        6 & 200 & 5  & 0.04 & 0.156 & 0.010 & 0.017 & 0.029 \\
        7 & 100 & 10 & 0.04 & 0.037 & 0.003 & 0.009 & 0.029 \\
        8 & 200 & 10 & 0.04 & 0.040 & 0.005 & 0.008 & 0.013 \\ \hline
    \end{tabular}

    \caption{Experiment 1, stationary anisotropic process, \textbf{informative priors} -- estimated relative bias (top part) and relative mean squared error (bottom part) for different model parameters (columns) and for different models considered in this experiment (rows).}
    \label{tab:Exp1_numeric_results_informative}
\end{table}

\begin{table}[t]
    \renewcommand{\arraystretch}{1.3}

    \centering
    \begin{tabular}{c|ccc|cccc}
        & \multicolumn{3}{c|}{Model parameters} & \multicolumn{3}{c}{Relative bias} \\ \hline
        Model & $\lambda$ & $\alpha$ & $\sigma$ & $\alpha$ & $\sigma_x$ & $\sigma_y$ \\ \hline
        1 & 100 & 5  & 0.02 & 0.080 & 0.003 & 0.042 \\
        2 & 200 & 5  & 0.02 & 0.004 & 0.044 & 0.017 \\
        3 & 100 & 10 & 0.02 & 0.032 & 0.021 & 0.042 \\
        4 & 200 & 10 & 0.02 & -0.019 & -0.001 & -0.007 \\
        5 & 100 & 5  & 0.04 & 0.098 & 0.026 & 0.070 \\
        6 & 200 & 5  & 0.04 & 0.309 & 0.157 & 0.171 \\
        7 & 100 & 10 & 0.04 & 0.102 & -0.001 & 0.038 \\
        8 & 200 & 10 & 0.04 & -0.027 & 0.023 & 0.044 \\ \hline
    \end{tabular}

    \vspace{0.2cm}

    \begin{tabular}{c|ccc|cccc}
        & \multicolumn{3}{c|}{Model parameters} & \multicolumn{3}{c}{Relative MSE} \\ \hline
        Model & $\lambda$ & $\alpha$ & $\sigma$ & $\alpha$ & $\sigma_x$ & $\sigma_y$ \\ \hline
        1 & 100 & 5  & 0.02 & 0.034 & 0.018 & 0.018 \\
        2 & 200 & 5  & 0.02 & 0.023 & 0.034 & 0.006 \\
        3 & 100 & 10 & 0.02 & 0.014 & 0.010 & 0.008 \\
        4 & 200 & 10 & 0.02 & 0.007 & 0.006 & 0.004 \\
        5 & 100 & 5  & 0.04 & 0.152 & 0.047 & 0.054 \\
        6 & 200 & 5  & 0.04 & 0.326 & 0.203 & 0.082 \\
        7 & 100 & 10 & 0.04 & 0.048 & 0.010 & 0.024 \\
        8 & 200 & 10 & 0.04 & 0.023 & 0.010 & 0.015 \\ \hline
    \end{tabular}

    \caption{Experiment 2, stationary isotropic process, \textbf{uniform priors} -- estimated relative bias (top part) and relative mean squared error (bottom part) for different model parameters (columns) and for different models considered in this experiment (rows).}
    \label{tab:Exp2_numeric_results_uniform}
\end{table}

\clearpage

\begin{table}[h]
    \renewcommand{\arraystretch}{1.3}

    \centering
    \begin{tabular}{c|cccc|ccccc}
        & \multicolumn{4}{c|}{Model parameters} & \multicolumn{5}{c}{Relative bias} \\ \hline
        Model & $\lambda$ & $\alpha$ & $\sigma$ & $\theta_1$ & $\alpha$ & $\sigma_x$ & $\sigma_y$ & $\theta_0$ & $\theta_1$ \\ \hline
        1 & 200 & 5  & 0.02 & 0.5 & 0.041 & 0.025 & 0.022 & 0.023 & -0.018 \\
        2 & 200 & 10 & 0.02 & 0.5 & 0.031 & 0.013 & 0.014 & 0.004 & -0.015 \\
        3 & 200 & 5  & 0.02 & 1   & 0.059 & -0.008 & 0.027 & -0.030 & 0.030 \\
        4 & 200 & 10 & 0.02 & 1   & 0.012 & 0.015 & 0.016 & 0.005 & 0.004 \\ \hline
    \end{tabular}

    \vspace{0.2cm}

    \begin{tabular}{c|cccc|ccccc}
        & \multicolumn{4}{c|}{Model parameters} & \multicolumn{5}{c}{Relative MSE} \\ \hline
        Model & $\lambda$ & $\alpha$ & $\sigma$ & $\theta_1$ & $\alpha$ & $\sigma_x$ & $\sigma_y$ & $\theta_0$ & $\theta_1$ \\ \hline
        1 & 200 & 5  & 0.02 & 0.5 & 0.014 & 0.005 & 0.006 & 0.010 & 0.007 \\
        2 & 200 & 10 & 0.02 & 0.5 & 0.009 & 0.003 & 0.007 & 0.005 & 0.009 \\
        3 & 200 & 5  & 0.02 & 1   & 0.019 & 0.004 & 0.013 & 0.022 & 0.018 \\
        4 & 200 & 10 & 0.02 & 1   & 0.006 & 0.007 & 0.004 & 0.011 & 0.005 \\ \hline
    \end{tabular}

    \caption{Experiment 3, non-stationary process with anisotropic clusters, \textbf{uniform priors} -- estimated relative bias (top part) and relative mean squared error (bottom part) for different model parameters (columns) and for different models considered in this experiment (rows).}
    \label{tab:Exp3_numeric_results_uniform}
\end{table}

\begin{table}[h]
    \renewcommand{\arraystretch}{1.3}

    \centering
    \begin{tabular}{c|cccc|ccccc}
        & \multicolumn{4}{c|}{Model parameters} & \multicolumn{5}{c}{Relative bias} \\ \hline
        Model & $\lambda$ & $\alpha$ & $\sigma_{0,x}$ & $\sigma_{1,x}$ & $\alpha$ & $\sigma_{0,x}$ & $\sigma_{1,x}$ & $\sigma_{0,y}$ & $\sigma_{1,y}$ \\ \hline
        1 & 200 & 5  & $\log(0.01)$ & 1   & 0.011 & 0.002 & 0.030 & 0.001 & 0.009 \\
        2 & 200 & 10 & $\log(0.01)$ & 1   & 0.017 & 0.002 & 0.002 & 0.004 & 0.019 \\
        3 & 200 & 5  & $\log(0.01)$ & 1.5 & 0.101 & -0.004 & 0.047 & 0.004 & 0.028 \\
        4 & 200 & 10 & $\log(0.01)$ & 1.5 & 0.002 & 0.006 & 0.027 & 0.004 & 0.026 \\ \hline
    \end{tabular}

    \vspace{0.2cm}

    \begin{tabular}{c|cccc|ccccc}
        & \multicolumn{4}{c|}{Model parameters} & \multicolumn{5}{c}{Relative MSE} \\ \hline
        Model & $\lambda$ & $\alpha$ & $\sigma_{0,x}$ & $\sigma_{1,x}$ & $\alpha$ & $\sigma_{0,x}$ & $\sigma_{1,x}$ & $\sigma_{0,y}$ & $\sigma_{1,y}$ \\ \hline
        1 & 200 & 5  & $\log(0.01)$ & 1   & 0.012 & 0.001 & 0.066 & 0.001 & 0.048 \\
        2 & 200 & 10 & $\log(0.01)$ & 1   & 0.007 & 0.000 & 0.036 & 0.001 & 0.047 \\
        3 & 200 & 5  & $\log(0.01)$ & 1.5 & 0.026 & 0.001 & 0.061 & 0.001 & 0.028 \\
        4 & 200 & 10 & $\log(0.01)$ & 1.5 & 0.012 & 0.001 & 0.035 & 0.001 & 0.063 \\ \hline
    \end{tabular}

    \caption{Experiment 4, non-stationary process with isotropic clusters, \textbf{uniform priors} -- estimated relative bias (top part) and relative mean squared error (bottom part) for different model parameters (columns) and for different models considered in this experiment (rows).}
    \label{tab:Exp4_numeric_results_uniform}
\end{table}

\clearpage

\section{Illustration of credible envelopes}\label{app:credible_envelope}

\begin{figure}[h]
    \centering
    \includegraphics[width=0.7\textwidth]{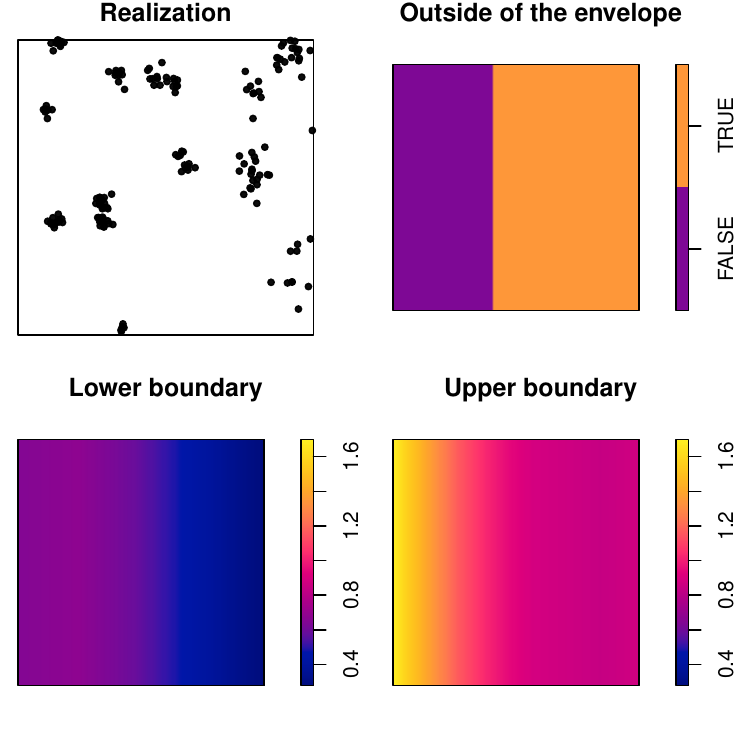}
    \caption{Illustration of the credible envelope.}
    \label{fig:credible_envelope}
\end{figure}

In this section we provide a graphical illustration of the use of the credible envelopes discussed in Section~\ref{sec:test_of_isotropy}. The illustration is given in Figure~\ref{fig:credible_envelope}.

The top left panel shows the point pattern used in this illustration, generated from model 4 in Experiment 4, see Section~\ref{sec:simulations} for details. The credible envelope is constructed as described in Section~\ref{sec:test_of_isotropy}. The bottom left and bottom right panels of Figure~\ref{fig:credible_envelope} show the resulting boundaries of the credible envelope.

Finally, we consider the constant function with value 1 and note that it lies outside of the credible envelope in the right part of the observation window, see the top right panel of Figure~\ref{fig:credible_envelope}. Hence, we reject the hypothesis of isotropic clusters. Note that this is a false positive result since the pattern was in fact simulated from a model with isotropic clusters.

Interpreting the graphical results is not straightforward. The top right panel of Figure~\ref{fig:credible_envelope} shows that the (estimated) diagonal template matrix $\Sigma_0(u)$ is not a multiple of the identity matrix in the right part of the observation window. Since the benchmark function lies above the upper boundary of the envelope in this region, we see that $\sigma_x(u) < \sigma_y(u)$ here. The rotation angle $\theta(u)$ now comes into play. The model fitted in Experiment~4 considered a constant angle $\theta(u) = \theta_0$. The values of $\theta_0$ sampled during the MCMC run were close to 0, indicating that little to no rotation is needed to describe the observed point pattern. Altogether, we conclude that the clusters are, according to the model fitted by the Bayesian MCMC procedure, elongated along the vertical axis.

\begin{figure}[h]
    \centering
    \includegraphics[width=\textwidth]{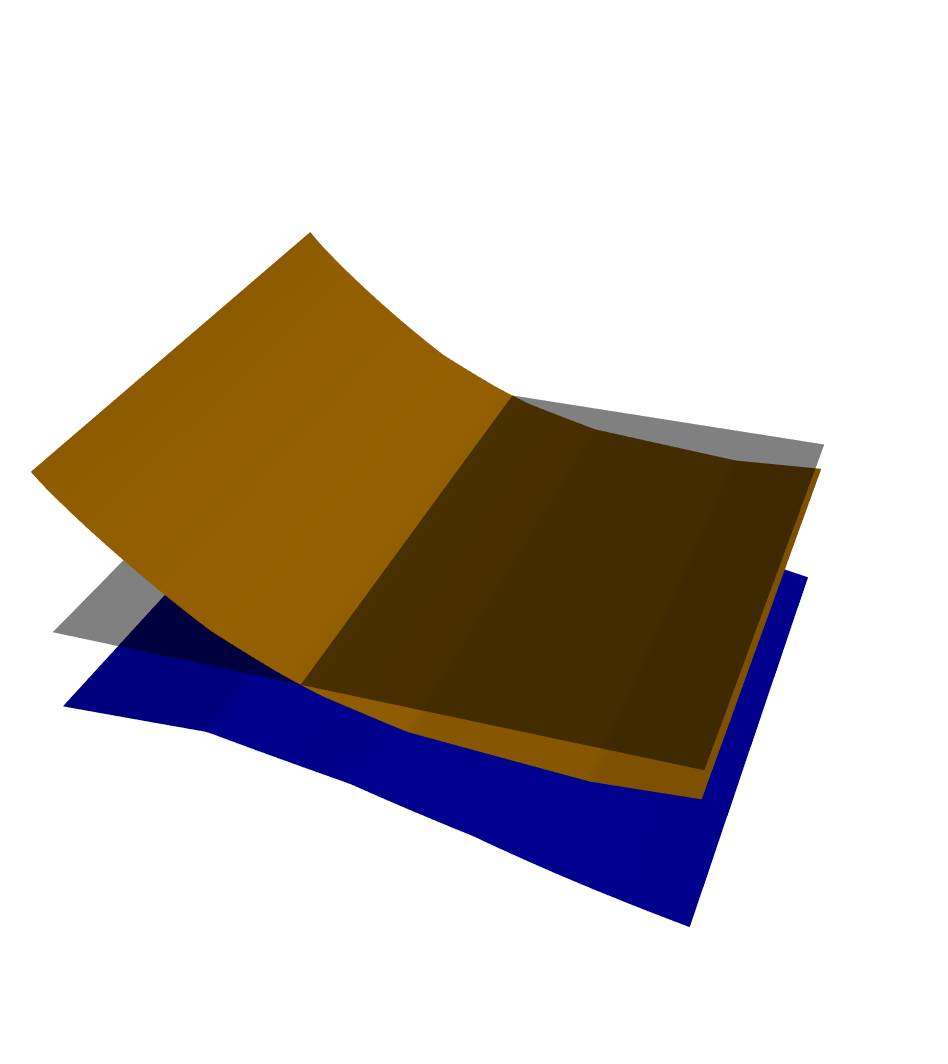}
    \caption{Illustration of the credible envelope in 3D.}
    \label{fig:credible_envelope3D}
\end{figure}

Figure~\ref{fig:credible_envelope3D} shows the credible envelope in three dimensions for better clarity. Note that the values of the upper and lower envelopes are shown in orange and blue colors, respectively, while the benchmark function 1 is shown in semi-transparent black.

\end{document}